\documentclass[journal=jacsat,manuscript=article]{achemso}

\usepackage{chemformula} 
\usepackage[T1]{fontenc} 
\usepackage{bm}
\usepackage{mathtools}
\usepackage[normalem]{ulem}

\newcommand{\Lp}{L_{\mathrm{p}}}
\newcommand{\kB}{k_{\mathrm{B}}}
\newcommand{\Rc}{\bm{R}_{\mathrm{c}}}
\newcommand{\dotRc}{\dot{\bm{R}}_{\mathrm{c}}}

\newcommand{\figref}[1]{Fig.\,\ref{#1}}

\newcommand{\figrefp}[2]{Fig.\,\ref{#1}\,(#2)}

\author{Zhongqiang Xiong}
\affiliation{Wenzhou Key Laboratory of Biomaterials and Engineering, Wenzhou Institute, University of Chinese Academy of Sciences, Wenzhou, 325000, China}
\alsoaffiliation{Institute of Theoretical Physics, Chinese Academy of Sciences, Beijing, 100190, China}

\author{Ryohei Seto}
\affiliation{Wenzhou Key Laboratory of Biomaterials and Engineering, Wenzhou Institute, University of Chinese Academy of Sciences, Wenzhou, 325000, China}
\alsoaffiliation{Oujiang Laboratory (Zhejiang Lab for Regenerative Medicine, Vision and Brain Health), Wenzhou, 325000, China}
\alsoaffiliation{Graduate School of Information Science, University of Hyogo, Kobe, 650-0047, Japan}

\author{Masao Doi}
\affiliation{Wenzhou Key Laboratory of Biomaterials and Engineering, Wenzhou Institute, University of Chinese Academy of Sciences, Wenzhou, 325000, China}
\alsoaffiliation{Oujiang Laboratory (Zhejiang Lab for Regenerative Medicine, Vision and Brain Health), Wenzhou, 325000, China}
\email{doi.masao@a.mbox.nagoya-u.ac.jp}

\title{Bending--Rotation coupling in the viscoelasticity of semiflexible polymers\\
---Rigorous perturbation analysis \\ from the rod limit}

\begin{document}
{\centering (Submitted: March 8, 2024. Accepted: April 19, 2024. doi: 10.1021/acs.macromol.4c00532)}
\bigskip
%
%
%
%
%

\begin{abstract}
  Brownian motion and viscoelasticity of semiflexible polymers is a subject that has been studied for many years.
  Still, rigorous analysis has been hindered due to the difficulty in handling the constraint that polymer chains cannot be stretched along the contour. 
  Here, we show a straightforward method to solve the problem. 
  We consider a stiff polymer that has a persistent length $\Lp$ much larger than the contour length $L$.
  We express the polymer configuration using three types of variables: the position vector of the center of mass $\Rc$, the unit vector $\bm{n}$ along the main axis, and the normal coordinates $\bm{u}_p$ for bending.
  Solving the Smoluchowski equation for the distribution function of these variables, we calculate the equilibrium time correlation 
  function $ \langle \bm{P}(t)\cdot \bm{P}(0) \rangle$ of the end-to-end vector $\bm{P}$ and the complex modulus $G^*(\omega)$ of dilute solution.
  They include the bending effect to the first order in $\theta \equiv L/\Lp$ and reduce to the exact results for the rigid rod in the limit of $\theta \to 0$. 
  The rotational diffusion coefficient increases slightly by the semiflexibility because the equilibrium length of the semiflexible polymer is smaller than that of the rigid rod with the same contour length.
  The storage modulus shows the same asymptotic dependence $G'(\omega) \sim \omega^{3/4}$ predicted by Shankar, Pasquali, and Morse [J. Rheol. 2002, 46, 1111--1154].
  The high-frequency viscosity is predicted to be dependent on the thickness of the semiflexible polymers.
\end{abstract}

\section{Introduction}

Semiflexible polymer is an important class of polymers that includes many biological systems, double-stranded DNA\cite{1994_Perkins}, F-actin\cite{1996_Kas,1996_Amblard,2002_Le-Goff}, cytoskeletons\cite{2006_Deng,2010_Fletcher}, microtubules\cite{1993_Gittes}, intermediate filaments\cite{1999_Everaers}, etc.
Such objects are usually modeled by a thin rod that can be bent but cannot be stretched along the contour.
The model is characterized by two lengths: the contour length $L$ and the persistent length $\Lp \equiv B/\kB T$, where $B$ is the bending modulus, $\kB$ is the Boltzmann constant, and $T$ is the temperature of the system.  
The investigation of semiflexibility started in 1949; \citet{1949_Kratky} proposed the worm-like chain (WLC) model, which describes the equilibrium properties of semiflexible polymers.
Calculation of the equilibrium quantities of this model is relatively easy, and many analytical results are known. 
For example, the mean square of the end-to-end vector $\bm{P}$ is given by\cite{1986_Doi}
\begin{equation}
\big\langle\bm{P}^2\big\rangle_{\mathrm{eq}}
= 2 \Lp^2 \left[
\exp \biggl( -\frac{L}{\Lp} \biggr)
+\frac{L}{\Lp}-1\right] 
=L^2\left(1 - \frac{1}{3}\theta \right)
+ o(\theta), \label{eqn:1}
\end{equation}
where the second equality is obtained for $\theta \equiv L/\Lp = \kB T L/B\ll 1$.


On the other hand, dynamical quantities of the semiflexible polymer are much more difficult to calculate.
There is no analytical expression for the time correlation function $\langle \bm{P}(t)\cdot \bm{P}(0) \rangle$
of the end-to-end vector $\bm{P}$.
The difficulty arises from the constraint that the polymer is inextensible along the contour.
The dynamics of semiflexible polymer is usually dealt with by the Langevin equation\cite{1998_Morse}.
In this treatment, the inextensible constraint is accounted for by introducing an unknown tensile force in the Langevin equation.
The tensile force must be calculated by solving another nonlinear partial differential equation.
%
Therefore, it becomes difficult to obtain analytical solution.

Earlier work of \citet{1966_Harris} and \citet{1966_Hearst} used a physical argument to simplify the problem and derived an expression for the complex modulus $G^*(\omega)$, but their result did not give the correct behavior of rigid rod, i.e., in the limit of $\theta \to 0$. 
\citet{1998_Gittes} constructed the Langevin equation for the tensile force, which accounts for the inextensibility of the polymer.
They showed the importance of the tensile force in $G^*(\omega)$ and derived a nontrivial scaling relation $G'(\omega) \sim \omega^{3/4}$, which was confirmed by experiments\cite{1996_Amblard}.
Many previous theories used the Fourier transform to calculate the bending energy of the polymer\cite{1993_Gittes,1997_Granek,1998_Gittes,2014_Broedersz}.
However, the Fourier decomposition is not compatible with the boundary conditions for the forth-order differential equation for bending and does not give accurate relaxation times\cite{1985_Aragon,1998_Wiggins,2007_Young,2012_Kantsler, 2017_De-Canio}. 

Constructing a more rigorous theory was attempted by Shankar, Pasquali, and Morse\cite{2002_Shankar}.
They conducted the analysis in the stiff polymer limit $\theta \ll 1$.
The inner tension along the polymer is introduced to satisfy the 
requirement of inextensibility. 
This gives an integro-differential equation for the tension. 
This work is perhaps the most elaborate and detailed analysis of the viscoelasticity of semiflexible polymer.
However, even in this work, they had to introduce an approximation (called local compliance approximation) to solve the integro-differential equation analytically, or the result was obtained by numerical calculations.


In this paper, we will take a different pathway to solve the problem. 
Instead of the Langevin equation, we will use the Smoluchowski equation, which does not explicitly consider the inner tensile force.
Up to the first order of $\theta$, we conduct the calculation following the standard procedure and obtain an analytical expression for the mean square displacement (MSD) of the center of mass, the end-to-end vector at equilibrium, and the complex modulus $G^*(\omega) $.  
In the limit of $\theta \to 0$, these expressions reduce to the exact results known for a rigid rod and confirm the validity of the present calculation. 

\section{Formulations}

We consider a semiflexible polymer having contour length $L$ and persistent length $\Lp$ placed in a solvent that is flowing with velocity gradient $\bm{\kappa}(t)$. 
We label each segment of the polymer using the contour coordinate $s$ along the polymer: $s$ changes from $-1$ to $1$ along the polymer. 
The position of segment $s$ at time $t$ is denoted by $\bm{R}(s;t)$. Therefore, $\bm{R}(s)$ denotes a conformation of the whole polymer chain.


One basic feature of semiflexible polymers is that the chain can store bending potential energy. 
The potential energy is written by the bending modulus $B$ and the curvature $(2/L)^2\bm{R}''$ of the chain\cite{1998_Gittes,1998_Morse}
\begin{equation}
    U=\int_{-1}^{1} ds\frac{1}{2}\tilde{B} \bm{R}''^2, \label{potential}
\end{equation}
under inextensibility condition\cite{1998_Wiggins}, where $\tilde{B}=(2/L)^3B$ and the double prime indicates the second-order derivative with respect to $s$.
%
%
The elastic force of semiflexible polymers is then given by the gradient of the functional $U$ with respect to $\bm{R}$, i.e.,
\begin{equation}
    \bm{F}_e = -\frac{\delta U}{\delta \bm{R}(s)} = -\tilde{B} \bm{R}''''(s), \label{elastic_force}
\end{equation}
where $\delta/\delta \bm{R}(s)$ denotes the functional derivative and the quadruple prime indicates the fourth-order derivative. 
Note that there is an additional term in eq.~\eqref{elastic_force} for unknown inner tensile force $\mu(s)$ (Lagrange multiplier) caused by the inextensibility. With both ends free, one has boundary conditions $\bm{R}''=\bm{R}'''=0$ and $\mu=0$ at $s=\pm 1$\cite{1998_Wiggins}.
However, we shall regard the inextensibility as geometric constraints later.
%
%
%
The motion of polymer segment is determined by the balance of the driving force (the force arising from the potential $U$) and the dissipative force (the frictional force exerted by surrounding fluids), and the governing Smoluchowski equation is obtained conveniently by the Onsager's Variational Principle\cite{1986_Doi}, which provides great flexibility in choosing the generalized coordinates. 
%
This is shown in the following.

Let $\psi[\bm{R}(s),t]$ be the probability of finding the polymer at the conformation $\bm{R}(s)$ at time $t$.
This satisfies the conservation equation
\begin{equation}
  \frac{\partial}{\partial t}\psi[\bm{R},t] = -\int_{-1}^{1} ds \frac{\delta}{\delta \bm{R}(s)} \cdot \big(\dot{\bm{R}}(s)\psi[\bm{R},t]\big). 
 \label{eq:1.0}
\end{equation}


The segment velocity $\dot{\bm{R}}(s)$ is obtained by minimizing a functional called Rayleighian $\mathcal{R}$\,\cite{1986_Doi}, which is a functional of $\dot{\bm{R}}(s)$ and is written as
\begin{equation}
	\mathcal{R}[\dot{\bm{R}}(s),\bm{\kappa}] = \Phi[\dot{\bm{R}}(s),\bm{\kappa}] + \dot{A}[\dot{\bm{R}}(s),\bm{\kappa}],\label{Rayleighian}
\end{equation}
where $\Phi$ is the dissipation function (the work done by the dissipative force per unit time) and $\dot{A}$ is the rate of the free energy change (i.e., the time derivative of the free energy).
If $\dot{\bm{R}}(s)$ is subject to certain constraints, they must be included in eq.\,\eqref{Rayleighian} with some Lagrange multipliers. 
$\dot{\bm{R}}(s)$ in eq.\,\eqref{eq:1.0} is given by the minimum condition of the Rayleighian, i.e.,
\begin{equation} \label{varitaional}
	\frac{\delta \mathcal{R}[\dot{\bm{R}}(s),\bm{\kappa}]}{\delta \dot{\bm{R}}(s)} = 0.
\end{equation}

Inserting the solution of eq.~\eqref{varitaional} for $\dot{\bm{R}}(s)$ into eq.~\eqref{eq:1.0}, one obtains the Smoluchowski equation of the system. 
Also using the solution for $\dot{\bm{R}}(s)$, the stress tensor $\bm{\sigma}$ of the system is calculated from the Rayleighian by\cite{1986_Doi}
\begin{equation}
    \bm{\sigma}[\bm{\kappa}] = 
    \frac{\partial \mathcal{R}[\dot{\bm{R}}(s),\bm{\kappa}]}{\partial \bm{\kappa}}. 
    \label{eq:stress}
\end{equation}

\subsection{Basic variables for conformations}

The translational diffusion of the center-of-mass, rotational diffusion of the main axis of the polymer, and the bending fluctuations perpendicular to the axis are three motions (modes) to characterize the dynamics of semiflexible polymers. 
%
To find proper coordinates to represent such modes, we use the following two steps. 

Firstly, we use the eigenfunctions $f_p(s)$ defined by the following eigenvalue equation corresponding to eq.~\eqref{elastic_force}
\cite{1985_Aragon, 1998_Wiggins,2002_Shankar, 2007_Young, 2012_Kantsler}  
\begin{equation}
    f_p''''(s)= \lambda_p^4 f_p(s), \label{equ:e1}
\end{equation}
and the boundary conditions
\begin{equation}
	  \left.f_p''(s)\right|_{s=\pm1} = \left.f_p'''(s)\right|_{s=\pm1}=0, \label{equ:e2}
\end{equation}
where $f_p'''(s)$ indicates the third-order derivative of $f_p(s)$.
%
%
The eigenfunctions can be taken to be orthonormal,
\begin{equation}
    \int_{-1}^1 ds f_p(s)f_q(s) = \delta_{pq}. \label{equ:e1A}
\end{equation}
The explicit form of the eigenvalues and the eigenfunctions are given in Appendix ``Eigen solutions.''
%
There are two eigenfunctions having zero eigenvalues, which correspond to the translation and the rotation of the polymer.
We shall denote these eigenfunctions with suffixes `$\text{trans}$' and $0$, respectively.
\begin{subequations}
\begin{gather}
 \lambda_{\text{trans}} = 0, \qquad f_{\text{trans}}(s)=\sqrt{1/2},     \label{eqn:2}  \\ 
    \lambda_{0} = 0, \qquad f_0(s)=\sqrt{3/2} s.   \label{eqn:3} 
\end{gather}
\end{subequations}
The other eigenfunctions denoted by the suffixes $p=1,2,3,\dots$ correspond to the bending of the polymer. The orthogonality of the eigenfunctions \eqref{equ:e1A}
gives the following relation 
\begin{equation}\label{equ:e3}
	\int_{-1}^{1} ds f_p(s) = 0, \qquad
		\int_{-1}^{1} ds f_p(s) s = 0,
\end{equation}
for $p=1,2,3,\dots$, i.e., the bending modes are orthogonal to the translational and rotational modes.

Using the eigenfunctions $f_p(s)$, we can express $\bm{R}(s)$ as
\begin{equation}
	\bm{R}(s)= \Rc+ \bm{N}s + \sum_{p=1}^\infty \bm{U}_pf_p(s). \label{equ:e4a}
\end{equation}
The orthogonality of the eigenfunctions \eqref{equ:e1A} allows us to 
invert eq.\,\eqref{equ:e4a} and express $\Rc$, $\bm{N}$, and $\bm{U}_p$ in terms of $\bm{R}(s)$:
\begin{subequations}\label{eq:10abc}
\begin{align}
   \Rc &= \frac{1}{2} \int_{-1}^{1} ds \bm{R}(s),
  \label{eq:10abc1}\\
   \bm{N} &= \frac{3}{2} \int_{-1}^{1} ds \bm{R}(s)s, 
     \label{eq:10abc2}\\   
   \bm{U}_p &= \int_{-1}^{1} ds \bm{R}(s)f_p(s).                         \label{eq:10abc3}
 \end{align}
\end{subequations}
Eq.~\eqref{eq:10abc1} indicates that $\Rc$ is the position vector of the center of mass, eq.~\eqref{eq:10abc2} indicates that $\bm{N} $ is a vector denoting the direction of the main axis, and eq.~\eqref{eq:10abc3} indicates that $\bm{U}_p$ ($p=1,2,3,\dots$) are the vector of the deflection from the straight conformation.  
Notice that the main axis vector $\bm{N}$ 
is uniquely expressed in terms of $\bm{R}(s)$ by such eigen mode analysis. Eq.~\eqref{equ:e4a} and eqs.~\eqref{eq:10abc1}--\eqref{eq:10abc3} represent a linear transformation for coordinates and are valid for any polymer conformation.  

Secondly, we consider that the polymer is inextensible and that $\bm{R}(s)$ is subject to the constraint
\begin{equation}
  \left(\frac{\partial \bm{R}}{\partial s}\right)^2 = 
   \left(\frac{L}{2}\right)^2,
   \label{eqn.13}
\end{equation}
for each point $s$. For such a polymer, the coordinates $\bm{N}$ and $\bm{U}_p$ have to satisfy certain constraints.  We now consider how to deal with such constraints.

%
It must be noted that although $\bm{R}(s)$ is close to that of a straight rod, the vector $\bm{N}$ defined by eq.\,\eqref{eq:10abc2} is subject to bending fluctuation, and its magnitude $|\bm{N}|$ is not constant.
To avoid this problem, we define the unit vector
\begin{subequations}
\begin{equation}
 \bm{n} \equiv \frac{\bm{N}}{|\bm{N}|},
\end{equation}
and decompose $\bm{U}_p$ into components, parallel and perpendicular to $\bm{n}$, 
i.e., $\bm{U}_p = \bm{U}_p^\parallel+ \bm{U}_p^\perp$. 
To simplify the notation, we denote the
perpendicular coordinate $\bm{U}_p^\perp$ by $\bm{u}_p$:
\begin{equation}
    \bm{u}_p \equiv \bm{U}_p^\perp = (\bm{\delta}-\bm{n}\bm{n})\cdot\bm{U}_p. \label{eqn.15}
\end{equation}
\end{subequations}
They have to satisfy the constraints
\begin{subequations}\label{eq:15.00}
	\begin{gather}
		\bm{n}\cdot\bm{n} = 1, \label{eq:15.0a}\\
		\bm{n}\cdot\bm{u}_p = 0.\label{eq:15.0b}
	\end{gather}
\end{subequations}

\begin{figure}[htpb]
\centering
\includegraphics[width=0.6\textwidth]{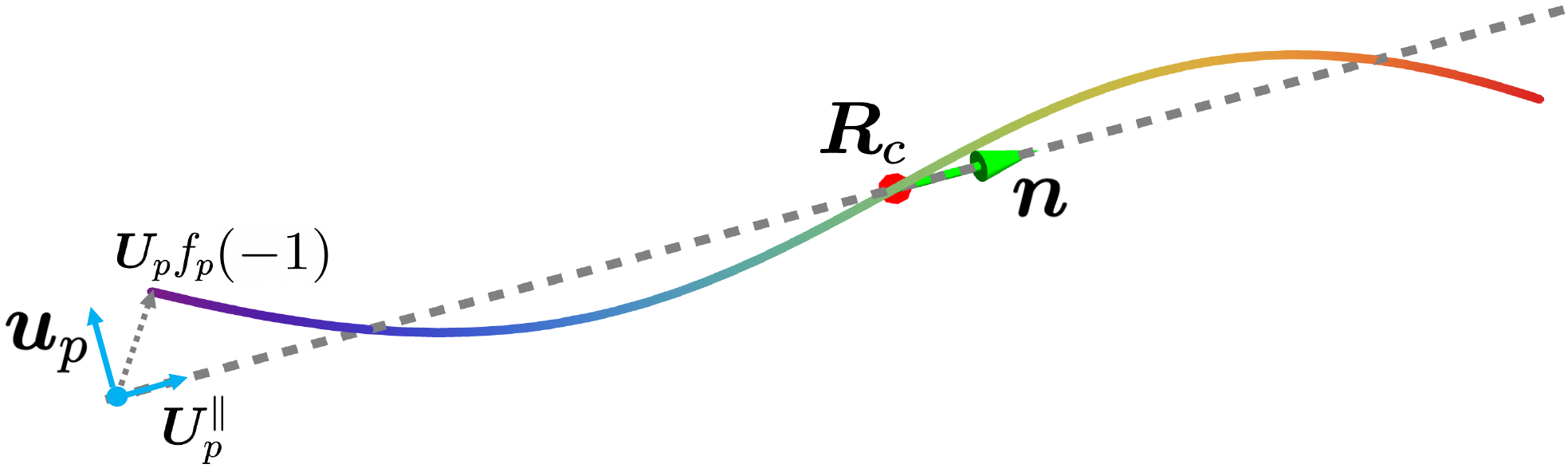}
\caption{\label{fig:graph}
An illustration for the coordinates introduced by eq.~\eqref{eq:10abc} with one bending mode, where the displacement of the segment at label $s=-1$ has been shown before (dashed line) and after (solid line) deformation. 
The color gradient (from purple to red)
indicates the coordinate $s$ from $-1$ to $1$.
$\bm{U}_p^\parallel$ exists because of the inextensibility of the semiflexible polymer.}
\end{figure}

It should be noted that for stiff polymer of $\theta \ll 1$,
$\bm{U}_p^\parallel$ is much smaller than $\bm{u}_p$. 
As it is shown in Appendix ``Inextensibility,'' in the stiff polymer limit of $\theta \ll 1$,  $\bm{U}_p^\parallel$ can be expressed in terms of $\bm{n}$ and $\bm{u}_p$ as follows:
\begin{subequations} \label{eq:un}
\begin{equation}
    \bm{U}_p^\parallel 
    = \bm{n}\left(\frac{2}{L} 
             \sum_{q=1}^\infty\sum_{r=1}^\infty \Gamma_{pqr}\bm{u}_q\cdot\bm{u}_r\right) + o(\theta),
  \label{eqn:16}
\end{equation}
where $\Gamma_{pqr}$ are constants defined in \eqref{eq:15a.0A}.
%
Also $\bm{N}$ can be expressed by $\bm{n}$ and $\bm{u}_p$ as
\begin{equation}
	\bm{N} = \bm{n} \left(\frac{L}{2} 
                - \frac{1}{L}\sum_{p=1}^\infty\sum_{q=1}^\infty \Gamma_{pq}\bm{u}_p\cdot\bm{u}_q\right) + o(\theta),  
                       \label{11.b}   
\end{equation}
\end{subequations}
where $\Gamma_{pq}$ are constants defined in \eqref{Cpq}. 
%
Since $\bm{U}_p$ and $\bm{N}$ are 
expressed by $\bm{u}_p$ and $\bm{n}$, 
therefore, we can express $\bm{R}(s)$ only in terms of $\Rc$, $\bm{n}$, and $\bm{u}_p$ $(p=1,2,3,\dots)$. 
It is important to note that $\Rc$, $\bm{n}$, and $\bm{u}_p$ $(p=1,2,3,\dots)$ are a complete set of variables representing the polymer conformation. We shall call these variables basic variables in this work, and an illustration for it is shown in \figref{fig:graph}.

Taking the time derivative of eq.~\eqref{eq:15.00}, we have
\begin{subequations}\label{eq10:constraint}
	\begin{gather}
	\bm{n}\cdot\dot{\bm{n}} = 0, \label{eq10:constraint1}\\
	\bm{n}\cdot\dot{\bm{u}}_p + \dot{\bm{n}}\cdot\bm{u}_p = 0.\label{eq10:constraint2}
\end{gather}
\end{subequations}
$\dot{\bm{n}}$ and $\dot{\bm{u}}_p$ have to satisfy these constraints. 

\subsection{Dynamic equations for the basic variables}
Since polymer conformation is expressed by $\Rc$, $\bm{n}$, and 
$\bm{u}_p$, we consider the distribution function $\psi(\Rc,\bm{n},\bm{u}_p,t)$, and obtain the Smoluchowski equation for it.
The distribution function satisfies the conservation equation
\begin{equation}
    \dot{\psi}=-\frac{\partial }{\partial \Rc}
    \cdot(\dotRc \psi) -\frac{\partial }{\partial \bm{n}}\cdot(\dot{\bm{n}} \psi) -\sum_{p=1}^\infty\frac{\partial }{\partial \bm{u}_p}\cdot(\dot{\bm{u}}_p \psi), \label{equ:e4}
\end{equation}
and the normalization condition $\int d\Omega \psi =1$, where $d\Omega$ is the volume element in the conformation space, $d\Omega=d\Rc d\bm{n} \Pi_{p=1}^\infty d\bm{u}_p$.

For the semiflexible polymer, the potential energy is eq.~\eqref{potential}. 
By using the conformation distribution function $\psi$, the free energy of the system is given by\cite{1986_Doi}
\begin{align} \label{free-energy}
A & =  \upsilon\int d\Omega \psi \left(\int_{-1}^{1} ds\frac{1}{2}\tilde{B} \bm{R}''^2+\kB T\ln{\psi} \right) \notag \\
& = \upsilon\int d\Omega \psi \left(\int_{-1}^{1} ds\frac{1}{2}\tilde{B} \sum_{p=1}^\infty \sum_{q=1}^\infty \bm{U}_p \cdot \bm{U}_q f_p''(s)f_q''(s)+\kB T\ln{\psi} \right) \notag \\
& =  \upsilon\int d\Omega \psi \left(\frac{1}{2}\tilde{B} \sum_{p=1}^\infty\lambda_p^4 \bm{u}_p^2+\kB T\ln{\psi} \right)+o(\theta),
\end{align}
where $\upsilon$ is the number density of polymers. 
The second equality is obtained by using eq.~\eqref{equ:e4a}.
The integral by parts over $s \in [-1,1]$ is used twice (see eq.~\eqref{f2f2}) to obtain the last equality based on the eigenvalue equations, \eqref{equ:e1} and \eqref{equ:e2}, and the orthogonality \eqref{equ:e1A}.

The distribution function at equilibrium is given by the minimum of $A$ with
respect to $\psi$ subject to the normalization condition.
This gives
\begin{equation}
   \psi_\text{eq} \propto 
   \exp\biggl(- \frac{\tilde B}{2\kB T} \sum_{p=1}^\infty\lambda_p^4 \bm{u}_p^2 
    \biggr).
\end{equation} 

The free energy change rate (the time derivative of the free energy \eqref{free-energy}) is then given by
\begin{align} \label{Adot}
	\dot{A}&=\upsilon\int d\Omega \dot{\psi} \left(\frac{1}{2} \tilde{B} \sum_{p=1}^\infty\lambda_p^4 \bm{u}_p^2+\kB T  \ln{\psi} +\kB T \right)+o(\theta) \notag\\
	&=\upsilon\int d\Omega \psi \left[\kB T\left(\dot{\Rc} \cdot\frac{\partial \ln{\psi}}{\partial \Rc}+\dot{\bm{n}} \cdot\frac{\partial \ln{\psi}}{\partial \bm{n}}+\sum_{p=1}^\infty\dot{\bm{u}}_p \cdot\frac{\partial \ln{\psi}}{\partial \bm{u}_p} \right)+\tilde{B}\sum_{p=1}^\infty\lambda_p^4\dot{\bm{u}}_p \cdot\bm{u}_p\right]+o(\theta),
\end{align}
where the conservation equation \eqref{equ:e4}, and the integration by parts over the conformation space $\Omega$ has been used.

Next, we calculate the dissipation function $\Phi$ following the procedure in ref\cite{1986_Doi}. 
The energy dissipation is caused by the motion of polymer segment relative to the surrounding solvent and is written as 
\begin{equation} \label{dissip0A}
  \Phi =\upsilon \int d\Omega \psi \int_{-1}^{1} ds \int_{-1}^{1} ds'
             \frac{1}{2}\bm{V}(s) \cdot \bm{\zeta}(s,s') \cdot \bm{V}(s'),
\end{equation}
where $\bm{\zeta}(s,s')$ is the friction tensor, which is generally a functional of $\bm{R}(s)$. $\bm{V}(s)$ is the velocity of the segment $s$ relative to the surrounding solvent, i.e., $\bm{V}(s) = \dot{\bm{R}}(s)- \bm{\kappa}\cdot\bm{R}(s)$.
For semiflexible polymers, the friction tensor can be approximated as $\bm{\zeta}(s,s') =\delta(s-s') [\zeta^\parallel\bm{n}\bm{n}+\zeta^\perp(\bm{\delta}-\bm{n}\bm{n})]$, where $\zeta^\parallel$ (and $\zeta^\perp$) are constants which represent the energy dissipation when the segment is moving parallel (and perpendicular) to the main axis. For the cylinder rod, we have $\zeta^\parallel=\pi \eta_s L /\ln(L/a)$ and $\zeta^\perp=2\zeta^\parallel$ based on the Kirkwood theory\cite{1986_Doi}, where $\eta_s$ is the viscosity of the solvent and $a$ is the diameter of the rod. 
In the following we shall write the friction tensor as $\bm{\zeta}(s,s')=\delta(s-s') (\zeta_1\bm{\delta}-\zeta_2\bm{n}\bm{n})$ with $\zeta_1=\zeta^\perp$ and $\zeta_2=\zeta^\perp-\zeta^\parallel$. 
Therefore, the dissipation function of the system is 
given by
\begin{equation} \label{dissip}
\Phi =\upsilon \int d\Omega \psi \int_{-1}^{1} ds\left\{\frac{1}{2}\zeta_1 (\dot{\bm{R}}- \bm{\kappa}\cdot\bm{R})^2-\frac{1}{2}\zeta_2 
 \left[ \bm{n}\cdot(\dot{\bm{R}}- \bm{\kappa}\cdot\bm{R}) \right]^2 \right\}.
\end{equation}
Inserting eq.~\eqref{equ:e4a} into eq.~\eqref{dissip}, and evaluating the integral and using the orthogonality \eqref{equ:e1A}, we 
can express the energy dissipation function as a quadratic function of 
$\dotRc$, $\dot{\bm{n}}$ and $\dot{\bm{u}}_p$. 
This is given in the terms involving the factors $\zeta_1$ and $\zeta_2$ in eq.~\eqref{equ:e13A}.


The Rayleighian \eqref{Rayleighian}
of the present system can be written as
\begin{equation} \label{Raylei0}
	\mathcal{R} =  \Phi(\dotRc,\dot{\bm{n}},\dot{\bm{u}}_p,t)
 +\dot{A}(\dotRc,\dot{\bm{n}},\dot{\bm{u}}_p,t)+\int d\Omega\mu_0 \bm{n}\cdot\dot{\bm{n}}+\sum_{p=1}^\infty\int d\Omega\mu_p \left(\bm{n}\cdot\dot{\bm{u}}_p+\dot{\bm{n}}\cdot\bm{u}_p\right),
\end{equation}
where the third and the fourth terms represent the constraints \eqref{eq10:constraint1} and \eqref{eq10:constraint2}. 
$\mu_0$ and $\mu_p$ are the Lagrange multipliers.
From eqs.~\eqref{Adot}, \eqref{dissip}, and \eqref{Raylei0}, we have
\begin{align} \label{equ:e13A}
\mathcal{R} ={}&
    \upsilon\int d\Omega \psi \left\{\frac{1}{2}\zeta_1 \left[2(\dotRc -\bm{\kappa}\cdot\Rc)^2+\frac{2\bm{N}^2}{3}(\dot{\bm{n}}-\bm{\kappa}\cdot\bm{n})^2+\frac{4}{3} \bm{n}\cdot\bm{\kappa}\cdot\bm{n}\sum_{p=1}^\infty\sum_{q=1}^\infty \Gamma_{pq}\dot{\bm{u}}_p\cdot\bm{u}_q\right.\right. \notag \\
     &
    \left.\left.\qquad\qquad {}
    + \sum_{p=1}^\infty (\dot{\bm{u}}_p-\bm{\kappa}\cdot\bm{u}_p)^2\right]-\frac{1}{2}\zeta_2 \Bigg[2\left(\bm{n}\cdot(\dotRc -\bm{\kappa}\cdot\Rc)\right)^2+ \frac{2\bm{N}^2}{3}(\bm{n}\cdot\bm{\kappa}\cdot\bm{n})^2\right.\Bigg.\notag \\
     &
    \Bigg.\left.\qquad\qquad {}
  + \frac{4}{3}\bm{n}\cdot\bm{\kappa}\cdot\bm{n}\sum_{p=1}^\infty\sum_{q=1}^\infty \Gamma_{pq}\dot{\bm{u}}_p\cdot\bm{u}_q+ \sum_{p=1}^\infty \Big(\bm{n}\cdot(\dot{\bm{u}}_p-\bm{\kappa}\cdot\bm{u}_p)\Big)^2\Bigg] \right. \notag \\
    &
    \left.\qquad\qquad {} 
    + \kB T\left(\dot{\Rc} \cdot\frac{\partial \ln{\psi}}{\partial \Rc}+\dot{\bm{n}} \cdot\frac{\partial \ln{\psi}}{\partial \bm{n}}+\sum_{p=1}^\infty\dot{\bm{u}}_p \cdot\frac{\partial \ln{\psi}}{\partial \bm{u}_p} \right)+\tilde{B}\sum_{p=1}^\infty\lambda_p^4\dot{\bm{u}}_p \cdot\bm{u}_p\right.\notag \\
    &
    \left.
    \qquad\qquad {}
    + \frac{\mu_0}{\upsilon\psi}\bm{n}\cdot\dot{\bm{n}}+\sum_{p=1}^\infty\frac{\mu_p}{\upsilon\psi}\left(\bm{n}\cdot\dot{\bm{u}}_p+\dot{\bm{n}}\cdot\bm{u}_p\right)\right\}+ o(\theta),
\end{align}
where $\bm{N}^2$ is also represented in terms of $\bm{u}_p$ by eq.~\eqref{11.b}.
%

The velocities $\dotRc$, $\dot{\bm{n}}$ and $\dot{\bm{u}}_p$
are obtained by minimizing the Rayleighian, i.e., 
\begin{subequations}
\begin{align}
    \frac{\delta \mathcal{R}}{\delta \dotRc} &= \upsilon\psi \left[2\left(\zeta_1 \bm{\delta}-\zeta_2 \bm{n}\bm{n}\right)\cdot(\dotRc -\bm{\kappa}\cdot\Rc)+\kB T\frac{\partial \ln{\psi}}{\partial \Rc}\right]=0, \label{equ:e12}\\
    \frac{\delta \mathcal{R}}{\delta \dot{\bm{n}}} &=\upsilon\psi \left[ \frac{2}{3}\zeta_1 \bm{N}^2(\dot{\bm{n}}-\bm{\kappa}\cdot\bm{n})+\kB T\frac{\partial \ln{\psi}}{\partial \bm{n}}+\frac{\mu_0}{\upsilon\psi} \bm{n}+\sum_{p=1}^\infty\frac{\mu_p}{\upsilon\psi}\bm{u}_p\right]=0, \label{equ:e13} \\
    \frac{\delta \mathcal{R}}{\delta \dot{\bm{u}}_p} &=\upsilon\psi \left[ \left(\zeta_1 \bm{\delta}-\zeta_2 \bm{n}\bm{n}\right)\cdot(\dot{\bm{u}}_p-\bm{\kappa}\cdot\bm{u}_p)+\kB T\frac{\partial \ln{\psi}}{\partial \bm{u}_p}+\tilde{B}\lambda_p^4\bm{u}_p+\frac{\mu_p}{\upsilon\psi} \bm{n}\right.\notag \\
    &\qquad\quad\left.+\frac{2}{3}\left(\zeta_1-\zeta_2\right)\bm{n}\cdot\bm{\kappa}\cdot\bm{n}\sum_{q=1}^\infty \Gamma_{pq}\bm{u}_q\right]=0. \label{equ:e14}
\end{align}
\end{subequations}
Eq.~\eqref{equ:e12} can be solved for $\dotRc$
by using the Sherman--Morrison formula\cite{1950_Sherman} for the inverse of $\zeta_1 \bm{\delta}-\zeta_2 \bm{n}\bm{n}$.
Eqs.~\eqref{equ:e13} and \eqref{equ:e14} can be solved for 
$\dot{\bm{n}}$ and $\dot{\bm{u}}_p$ together with
eqs.~\eqref{eq:15.00} and \eqref{eq10:constraint}.
In this calculation we have replaced $\left(\bm{\delta}+c\sum_{p=1}^\infty\bm{u}_p\bm{u}_p\right)^{-1}$ with $\bm{\delta}-c\sum_{p=1}^\infty\bm{u}_p\bm{u}_p$ since the error is of the order of $\theta$. 
As a consequence, we obtain the following expressions for the velocities, which are correct up to the first order of $\theta$:
\begin{subequations}\label{17.0}
\begin{align}
    \dot{\Rc}={}&-D\left(\bm{\delta}+\frac{\zeta^\perp-\zeta^\parallel}{\zeta^\parallel} \bm{n}\bm{n}\right)\cdot\frac{\partial \ln{\psi}}{\partial \Rc}+\bm{\kappa}\cdot\Rc, \\
    \dot{\bm{n}}={}&-\left(\bm{\delta}-\bm{n}\bm{n}\right)\cdot\left(\frac{3D}{\bm{N}^2}\frac{\partial \ln{\psi}}{\partial \bm{n}} -\bm{\kappa}\cdot\bm{n}\right)-\frac{3}{2\bm{N}^2}\frac{\zeta^\parallel}{\zeta^\perp}\bm{n}\cdot\left(\bm{\kappa}+\bm{\kappa}^T\right)\cdot\sum_{p=1}^\infty\bm{u}_p\bm{u}_p, \\
    \dot{\bm{u}}_p ={}&-\left(\bm{\delta}-\bm{n}\bm{n}\right)\cdot\left(2D\frac{\partial \ln{\psi}}{\partial \bm{u}_p}-\bm{\kappa}\cdot\bm{u}_p\right)-\frac{1}{\tau_p}\bm{u}_p-\frac{2}{3}\frac{\zeta^\parallel}{\zeta^\perp}\bm{n}\cdot\bm{\kappa}\cdot\bm{n}\sum_{q=1}^\infty \Gamma_{pq}\bm{u}_q-\bm{u}_p\cdot\bm{\kappa}\cdot\bm{n}\bm{n}, \label{solution_up}
\end{align}
\end{subequations}
where $D=\kB T/2\zeta^\perp$ and $\tau_p=(L/2)^3\zeta^\perp/B\lambda_p^4$ with $p=1,2,3,\dots$. 
One can check that eq.~\eqref{17.0} satisfies the constraints \eqref{eq10:constraint} up to the order of $\theta$. 
Eq.~\eqref{17.0} show that the dynamics of rotation and bending are coupled with each other. 

By combining eq.~\eqref{equ:e4} and eq.~\eqref{17.0}, 
we can now write down the Smoluchowski equation including the three modes, i.e., translational mode, rotational mode, and the bending fluctuations modes:
\begin{equation} \label{Smoluchowski equation}
	\frac{\partial}{\partial t}\psi = \mathcal{L}\psi,
\end{equation}
where the differential operator $\mathcal{L}$ is defined by
\begin{align} \label{eq52.0}
	\mathcal{L}&=\frac{\partial }{\partial \Rc}\cdot\left[D\left(\bm{\delta}+\frac{\zeta^\perp-\zeta^\parallel}{\zeta^\parallel} \bm{n}\bm{n}\right)\cdot\frac{\partial}{\partial \Rc}-\bm{\kappa}\cdot\Rc\right] \notag \\
	 & \quad  +\frac{\partial }{\partial \bm{n}}\cdot\left[\left(\bm{\delta}-\bm{n}\bm{n}\right)\cdot\left(\frac{3D}{\bm{N}^2}\frac{\partial }{\partial \bm{n}} -\bm{\kappa}\cdot\bm{n}\right)+\frac{3}{2\bm{N}^2}\frac{\zeta^\parallel}{\zeta^\perp}\bm{n}\cdot\left(\bm{\kappa}+\bm{\kappa}^T\right)\cdot\sum_{p=1}^\infty\bm{u}_p\bm{u}_p\right]\notag \\
	& \quad +\sum_{p=1}^\infty\frac{\partial }{\partial \bm{u}_p}\cdot\left[\left(\bm{\delta}-\bm{n}\bm{n}\right)\cdot\left(2D\frac{\partial }{\partial \bm{u}_p}-\bm{\kappa}\cdot\bm{u}_p\right)+\frac{1}{\tau_p}\bm{u}_p + \frac{2}{3}\frac{\zeta^\parallel}{\zeta^\perp}\bm{n}\cdot\bm{\kappa}\cdot\bm{n}\sum_{q=1}^\infty \Gamma_{pq}\bm{u}_q+\bm{u}_p\cdot\bm{\kappa}\cdot\bm{n}\bm{n}\right]. 
\end{align}

Eq.~\eqref{Smoluchowski equation} describes the evolution of the conformational distribution of the polymer.  
The equation is the base of calculating the correlation functions at equilibrium and also the linear response of the system to external perturbations. However, 
we do not need to solve eq.~\eqref{Smoluchowski equation} explicitly. These quantities can be calculated as shown in the following. 

We define the Green function 
$\mathcal{G}(\Omega,t; \Omega',t')$ which satisfies\,\cite{1986_Doi}
\begin{equation}\label{equ:e43}
   \frac{\partial }{\partial t}\mathcal{G}=\mathcal{L} \mathcal{G}
   \quad\text{with}\quad 
   \mathcal{G}|_{t=t'}=\delta(\Omega - \Omega').
\end{equation}

The Green function is the conditional probability that the polymer is in the conformation $\Omega$ at time $t$, given that it was in the conformation $\Omega'$ at time $t'$.
Consider for example to calculate the correlation
of two physical quantities $\bm{x}(t)$ at time $t$
and $\bm{y}(t')$ at time $t'$ for the system at 
equilibrium, where $\bm{x}(t)$ (and $\bm{y}(t)$) can be expressed as a function of $\Rc(t)$, $\bm{n}(t)$ or $\bm{u}_p(t)$ ($p=1,2,3,\dots$).
In shorthand, $\bm{x}(t)=\tilde{\bm{x}}(\Omega)$ and $\bm{y}(t')=\tilde{\bm{y}}(\Omega')$.
Then, the time correlation for the 
two quantities $\bm{x}(t)$ and $\bm{y}(t')$ is calculated by\cite{1986_Doi}
\begin{equation}
  \big\langle \bm{x}(t)\bm{y}(t') \big\rangle 
 = \int d\Omega \int  d\Omega' \psi_\text{eq}(\Omega') 
            \mathcal{G}(\Omega,t;\Omega',t') \tilde{\bm{x}}(\Omega)\tilde{\bm{y}}(\Omega'),
\end{equation}
with $d\Omega=d\Rc(t) d\bm{n}(t) \Pi_{p=1}^\infty d\bm{u}_p(t)$ and $d\Omega'=d\Rc(t') d\bm{n}(t') \Pi_{p=1}^\infty d\bm{u}_p(t')$, and we have the normalization condition $\langle 1 \rangle = \int d\Omega \int  d\Omega' \psi_\text{eq} \mathcal{G} = \int d\Omega \psi = 1$.

The time derivative of the correlation function $\big\langle \bm{x}(t)\bm{y}(t') \big\rangle$ is then written as\cite{1986_Doi}
\begin{align}
	\frac{\partial}{\partial t}
 \big\langle \bm{x}(t)\bm{y}(t') \big\rangle 
 &= \int d\Omega \int  d\Omega' \psi_\text{eq} \tilde{\bm{x}}(\Omega)\tilde{\bm{y}}(\Omega') \mathcal{L} \mathcal{G}=\int d\Omega \int  d\Omega' \psi_\text{eq} \mathcal{G}\mathcal{L}^\dag \tilde{\bm{x}}(\Omega)\tilde{\bm{y}}(\Omega') \notag \\
  & =\big\langle\mathcal{L}^\dag \tilde{\bm{x}}(\Omega)\tilde{\bm{y}}(\Omega')\big\rangle, \label{equ:e48.0}
\end{align}
where the conjugate operator $\mathcal{L}^\dag$ is defined by
\begin{align}
    \mathcal{L}^\dag &= \left[D\frac{\partial}{\partial \Rc}\cdot \left(\bm{\delta}+\frac{\zeta^\perp-\zeta^\parallel}{\zeta^\parallel} \bm{n}\bm{n}\right)+\Rc\cdot\bm{\kappa}^T\right]\cdot\frac{\partial }{\partial \Rc} \notag \\
	&\quad +\left(\frac{3D}{\bm{N}^2}\frac{\partial }{\partial \bm{n}}+\bm{n}\cdot\bm{\kappa}^T\right)\cdot \left(\bm{\delta}-\bm{n}\bm{n}\right)\cdot\frac{\partial }{\partial \bm{n}} \notag \\
	&\quad +\sum_{p=1}^\infty\left[\left(2D\frac{\partial}{\partial \bm{u}_p}+\bm{u}_p\cdot\bm{\kappa}^T\right)\cdot\left(\bm{\delta}-\bm{n}\bm{n}\right)-\frac{1}{\tau_p}\bm{u}_p-\frac{2}{3}\frac{\zeta^\parallel}{\zeta^\perp}\bm{n}\cdot\bm{\kappa}\cdot\bm{n}\sum_{q=1}^\infty \Gamma_{pq}\bm{u}_q\right]\cdot\frac{\partial }{\partial \bm{u}_p}. \label{equ:L}
\end{align}
In eq.~\eqref{equ:e48.0}, the first equality is obtained by using eq.~\eqref{equ:e43}. 
The second equality is obtained by taking the integral by parts over the conformation space. 
Eqs.~\eqref{eq:65pd} and \eqref{eq:66pd} have been used to obtain the conjugate of the operator $\mathcal{L}$. 
In the following, we shall conduct the calculation using eq.~\eqref{equ:e48.0}.

\section{Brownian motion at equilibrium}

We first study the characteristic features of the Brownian motion of the semiflexible polymer at equilibrium.
Since $\bm{\kappa}$ is zero in this case, eq.~\eqref{equ:L} can be simplified as
\begin{align} \label{eqkis0}
    \mathcal{L}^\dag_0 \equiv \left.\mathcal{L}^\dag\right|_{\bm{\kappa}=0} &={} D\frac{\partial}{\partial \Rc}\cdot \left(\bm{\delta}+\frac{\zeta^\perp-\zeta^\parallel}{\zeta^\parallel} \bm{n}\bm{n}\right)\cdot\frac{\partial }{\partial \Rc} +\frac{3D}{\bm{N}^2}\frac{\partial }{\partial \bm{n}}\cdot \left(\bm{\delta}-\bm{n}\bm{n}\right)\cdot\frac{\partial }{\partial \bm{n}} \notag \\
    &\quad
    +\sum_{p=1}^\infty\left(2D\frac{\partial}{\partial \bm{u}_p}\cdot\left(\bm{\delta}-\bm{n}\bm{n}\right)-\frac{1}{\tau_p}\bm{u}_p\right)\cdot\frac{\partial }{\partial \bm{u}_p}.
\end{align}

We will study three topics: the translational diffusion of the center-of-mass, the rotational diffusion of the main axis, and the bending fluctuations.

\paragraph{(1) Translational diffusion of the center-of-mass}
We first calculate the mean square displacement (MSD) of the center-of-mass
in time $t$.
%
By eq.~\eqref{eqkis0}, the time derivative of the MSD of the center-of-mass satisfies
\begin{equation}
	\frac{\partial}{\partial t}
 \big\langle [\Rc(t)-\Rc(0)]^2 \big\rangle =
 \big\langle\mathcal{L}^\dag_0 [\Rc(t)-\Rc(0)]^2
 \big\rangle 
 = 2\left(3+\frac{\zeta^\perp-\zeta^\parallel}{\zeta^\parallel}\right)D.
\end{equation}
Therefore, we have
\begin{equation}
	\big\langle [\Rc(t)-\Rc(0)]^2 \big\rangle = 2\left(3+\frac{\zeta^\perp-\zeta^\parallel}{\zeta^\parallel}\right) D t.
\end{equation}
By using $\zeta^\perp=2\zeta^\parallel$, one has the diffusion constant of the center-of-mass
\begin{equation}
	D_{\mathrm{c}} \equiv \lim_{t \to \infty}\frac{1}{6t} \big\langle [\Rc(t)-\Rc(0)]^2 \big\rangle = \frac{4}{3}D = \frac{\kB T\ln(L/a)}{3\pi \eta_s L}.
\end{equation} 
The results agree with that of the rigid rod \cite{1986_Doi} and show that the bending fluctuations do not play an explicit role in the diffusion of the center-of-mass.

\paragraph{(2) Bending deflections}

The bending deflections $\bm{u}_p$ are the main signature indicating the difference between semiflexible polymers and rigid-rod polymers. 
Here, we calculate the time correlation 
of the bending deflections. 

By using eq.~\eqref{eqkis0}, the time evolutions of the correlations for the bending fluctuations are calculated as
\begin{subequations}
\begin{align}
    \frac{\partial}{\partial t}\langle \bm{u}_p(t)\cdot\bm{u}_q(0) \rangle &=
    \big\langle\mathcal{L}^\dag_0 \left[\bm{u}_p(t)\cdot\bm{u}_q(0)\right]
    \big\rangle
    =-\frac{1}{\tau_p}\langle \bm{u}_p(t)\cdot\bm{u}_q(0) \rangle, \label{utu0}\\
    \frac{\partial}{\partial t}\langle \bm{u}_p(t)\cdot\bm{u}_q(t) \rangle 
    &= \big\langle\mathcal{L}^\dag_0 \left[\bm{u}_p(t)\cdot\bm{u}_q(t)\right]
    \big\rangle = 
    8D\delta_{pq}-\left(\frac{1}{\tau_p}+\frac{1}{\tau_q}\right)\langle \bm{u}_p(t)\cdot\bm{u}_q(t) \rangle. \label{utut}
\end{align}
\end{subequations}
From eq.~\eqref{utut}, one can easily obtain the equilibrium values for $\big\langle \bm{u}_p^2 \big\rangle_\text{eq}$, which is
\begin{equation}
	\big\langle \bm{u}_p^2 \big\rangle_\text{eq} = 4D\tau_p = \frac{\kB T L^3}{4B\lambda_p^4}\sim \theta. \label{eqvalue}
\end{equation}
Eq.~\eqref{utu0} gives the solution
\begin{equation}
    \langle \bm{u}_p(t)\cdot\bm{u}_q(0) \rangle = 4D\tau_p\delta_{pq}
    \exp\biggl(-\frac{t}{\tau_p}
    \biggr). \label{equ:e42}
\end{equation}

\paragraph{(3) Rotational diffusion}

To consider the rotational relaxation, we calculate
$\langle \bm{n}(t)\cdot\bm{n}(0) \rangle $.
Again, by using eq.~\eqref{eqkis0}, the time evolution of the correlation for the rotation is given by
\begin{equation}
    \frac{\partial}{\partial t}\langle \bm{n}(t)\cdot\bm{n}(0) \rangle
    = 
    \big\langle\mathcal{L}^\dag_0 \left[\bm{n}(t)\cdot\bm{n}(0)\right]\big\rangle 
    = -6D 
    \left\langle\frac{1}{\bm{N}^2} \bm{n}(t)\cdot\bm{n}(0) 
    \right\rangle.
    \label{eq:orintation}
\end{equation}
Since $\bm{N}^2$ is written in terms of $\bm{u}_p$ by eq.~\eqref{11.b}, the rotation of the semiflexible polymer is affected by bending fluctuations. 
To solve eq.~\eqref{eq:orintation}, we insert eq.~\eqref{11.b} into eq.~\eqref{eq:orintation} to obtain
\begin{equation} \label{eq:orint1}
    \frac{\partial}{\partial t}\langle \bm{n}(t)\cdot\bm{n}(0) \rangle =-6D\left(\frac{2}{L}\right)^2 \langle\bm{n}(t)\cdot\bm{n}(0)\rangle-6D\left(\frac{2}{L}\right)^4\sum_{p=1}^\infty\sum_{q=1}^\infty \Gamma_{pq}\langle\bm{u}_p(t)\cdot\bm{u}_q(t) \bm{n}(t)\cdot\bm{n}(0) \rangle +o(\theta).
\end{equation}
Similarly, the time evolution of the last term is calculated as
\begin{align}
	\label{eq:orint2}
    \frac{\partial}{\partial t}\langle\bm{u}_p(t)\cdot\bm{u}_q(t) \bm{n}(t)\cdot\bm{n}(0) \rangle &= 
    \big\langle\mathcal{L}^\dag_0 \left[\bm{u}_p(t)\cdot\bm{u}_q(t) \bm{n}(t)\cdot\bm{n}(0)\right]\big\rangle \notag \\%
    &= - \left[6D\left(\frac{2}{L}\right)^2+\frac{1}{\tau_p}+\frac{1}{\tau_q}\right]\langle\bm{u}_p(t)\cdot\bm{u}_q(t) \bm{n}(t)\cdot\bm{n}(0) \rangle \notag \\
    &\quad +8D\delta_{pq}\langle\bm{n}(t)\cdot\bm{n}(0) \rangle +o(\theta),
\end{align}
where $p,q=1,2,3,\dots$. 
Eq.~\eqref{eq:orint1} and~\eqref{eq:orint2} give a closed equation set. 
We solve the equation using a Laplace transform
\begin{equation}
	\mathcal{T}[\langle \bm{n}(t)\cdot\bm{n}(0) \rangle]=\int_{0}^\infty dt \exp(-\varsigma t) \langle \bm{n}(t)\cdot\bm{n}(0) \rangle.
\end{equation}
For eq.~\eqref{eq:orint1}, one has
\begin{equation} \label{eq:laplac1}
	\left[\varsigma+6D\left(\frac{2}{L}\right)^2\right] \mathcal{T}[\langle \bm{n}(t)\cdot\bm{n}(0) \rangle]
 = 1-6D\left(\frac{2}{L}\right)^4 \sum_{p=1}^\infty\sum_{q=1}^\infty
 \Gamma_{pq}\mathcal{T}[\langle\bm{u}_p(t)\cdot\bm{u}_q(t) \bm{n}(t)\cdot\bm{n}(0) \rangle]+o(\theta),
\end{equation}
where the initial value $\langle\bm{n}(0)\cdot\bm{n}(0) \rangle=1$ has been used. 
Making a Laplace transform for eq.~\eqref{eq:orint2}, we have
\begin{align}
 \mathcal{T}[\langle\bm{u}_p(t)\cdot\bm{u}_q(t) \bm{n}(t)\cdot\bm{n}(0) \rangle]
 &=\delta_{pq}\frac{4D\tau_p+8D\mathcal{T}[\langle \bm{n}(t)\cdot\bm{n}(0) \rangle]}{\varsigma+6D\big(\left(2/L\right)^2+1/3D\tau_p\big)}+o(\theta),
\end{align}
where initial values 
$\langle\bm{u}_p(0)\cdot\bm{u}_q(0) \bm{n}(0)\cdot\bm{n}(0) \rangle=\langle\bm{u}_p\cdot\bm{u}_q\rangle_\text{eq}=4D\tau_p\delta_{pq}$ have been used (see eq.~\eqref{equ:e42}). Noting that $D\tau_p$ is quite small (see eq.~\eqref{eqvalue}), we have
\begin{align}
 \label{eq:laplac2}
 \mathcal{T}[\langle\bm{u}_p(t)\cdot\bm{u}_q(t) \bm{n}(t)\cdot\bm{n}(0) \rangle]
&= 4D\tau_p\delta_{pq} \mathcal{T}[\langle \bm{n}(t)\cdot\bm{n}(0) \rangle]+o(\theta) \notag \\
&=\langle\bm{u}_p\cdot\bm{u}_q\rangle_\text{eq}\mathcal{T}[\langle \bm{n}(t)\cdot\bm{n}(0) \rangle]+o(\theta),
\end{align}
which holds for finite $\varsigma$. 
Eq.~\eqref{eq:laplac2} indicates that up to order $\theta$, the bending relaxation is fast, and we can use the following relation:
\begin{equation} \label{decoupled}
	\langle\bm{u}_p(t)\cdot\bm{u}_q(t) \bm{n}(t)\cdot\bm{n}(0) \rangle=\langle\bm{u}_p\cdot\bm{u}_q\rangle_\text{eq}\langle \bm{n}(t)\cdot\bm{n}(0) \rangle+o(\theta).
\end{equation}

Inserting eq.~\eqref{eq:laplac2} into eq.~\eqref{eq:laplac1}, we have 
\begin{equation} \label{equ79}
	\left[\varsigma+6D\left(\frac{2}{L}\right)^2+6D\left(\frac{2}{L}\right)^4\sum_{p=1}^\infty\sum_{q=1}^\infty \Gamma_{pq}\langle\bm{u}_p\cdot\bm{u}_q\rangle_\text{eq}\right] \mathcal{T}[\langle \bm{n}(t)\cdot\bm{n}(0) \rangle]=1+o(\theta).
\end{equation}
From eq.~\eqref{11.b} (or \eqref{eq:16.0})
\begin{equation}
	\langle\bm{N}^2\rangle_\text{eq} = \left(\frac{L}{2}\right)^2 -  \sum_{p=1}^\infty\sum_{q=1}^\infty \Gamma_{pq}\langle\bm{u}_p\cdot\bm{u}_q\rangle_\text{eq}+o(\theta).
\end{equation}
Therefore, eq.~\eqref{equ79} becomes
\begin{equation} \label{eq:solution}
	\mathcal{T}[\langle \bm{n}(t)\cdot\bm{n}(0) \rangle]=\frac{1}{\varsigma+6D/\langle\bm{N}^2\rangle_\text{eq}}+o(\theta).
\end{equation}
The inverse Laplace transform of eq.~\eqref{eq:solution} gives
\begin{equation}
    \langle \bm{n}(t)\cdot\bm{n}(0) \rangle = 
    \exp \bigl(-2D_{\mathrm{r}}t \bigr), \label{eq:solnt}
\end{equation}
where the rotational diffusion coefficient is
\begin{equation}
 D_{\mathrm{r}}=\frac{3D}{\langle\bm{N}^2\rangle_\text{eq}}
 =D_\text{rod}\left(1 + \alpha_D\theta \right)+ o\left(\theta\right) ,\label{DiffsionC}
\end{equation}
with $D_\text{rod}=6\kB T/\zeta^\perp L^2=3\kB T\ln(L/a)/\pi\eta_s L^3$ for the rigid-rod limit\cite{1986_Doi}, and $\alpha_D \approx 0.25714$ is given by eq.~\eqref{alphaD} in Appendix ``Eigen solutions.''
Eq.~\eqref{DiffsionC} indicates that the rotational 
diffusion coefficient of semiflexible polymer is slightly 
larger than that of a rigid rod due to bending fluctuation.
We conjecture that this is because the equilibrium length of the semiflexible polymer is slightly smaller than that of the rigid rod with the same contour length. Eq.~\eqref{DiffsionC} can be checked by Brownian dynamics simulation, but we leave it for future work.
This is an example of the effect of bending--rotation coupling.

\subsection{Brownian motion of the end-to-end vector}

Once we have the explicit form of the time correlation for $\bm{n}$ and $\bm{u}_p$, we can calculate the time correlation function of the end-to-end vector $\bm{P}(t) \equiv \bm{R}(1,t)-\bm{R}(-1,t)$. 
By use of eq.~\eqref{equ:e4a}, this is written as
\begin{equation}
  \bm{P}(t) = 2\bm{N}(t)+2\sum_{p=2,\text{even}}^\infty f_p(1)\bm{U}_p(t).
\end{equation}
Hence, the mean square of $\bm{P}(t)$ is calculated as
\begin{align}
	\langle \bm{P}^2(t) \rangle &=4\langle \bm{N}^2(t)\rangle +8\sum_{p=2,\text{even}}^\infty f_p(1)\langle\bm{N}(t)\cdot\bm{U}_p(t)\rangle+4\sum_{p=2,\text{even}}^\infty f_p^2(1)\langle \bm{U}_p^2(t)\rangle \notag\\
    &=L^2 - 2\sum_{p=1}^\infty \int_{-1}^1ds f_p'^2(s)\langle\bm{u}_p^2(t)\rangle+8\sum_{p=2,\text{even}}^\infty \langle\bm{u}_p^2(t)\rangle +o(\theta) \notag \\
    &=L^2\left(1 - \alpha_P\theta\right) + o\left(\theta\right),   \label{eq:44.44}
\end{align}
where eq.~\eqref{eq:un} has been used to obtain the second equality. 
The third equality is obtained by using eq.~\eqref{eqvalue}, where $\alpha_P \approx 0.33333$ is given by eq.~\eqref{alphaP} in Appendix ``Eigen solutions''.
The mean square end-to-end vector of semiflexible polymer is smaller than $L^2$ because there is the inextensibility of polymers with the bending fluctuations.
The result agrees with that of equilibrium theory\cite{1949_Kratky,1986_Doi}, see also eq.~\eqref{eqn:1}.

\begin{figure}[htb]
\centering
\includegraphics[width=0.9\textwidth]{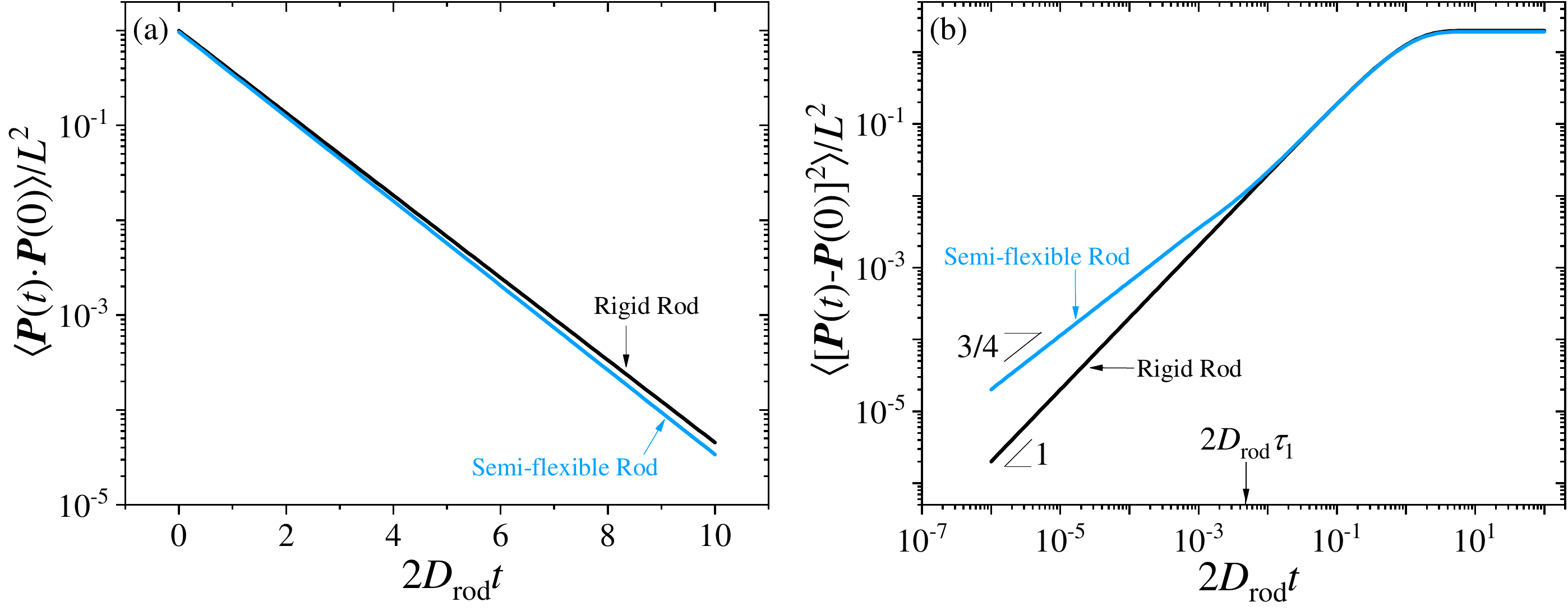}
\caption{\label{fig:f2.0}
Typical behaviors of (a) $\langle \bm{P}(t)\cdot\bm{P}(0) \rangle$ and (b) $\big\langle [\bm{P}(t)-\bm{P}(0)]^2 \big\rangle$ for $\theta=1/10$, where $D_\text{rod}$ is the rotational diffusion coefficient of the rigid-rod limit, and $\tau_1=\left(3/4\lambda_1^4\right)\theta/D_\text{rod}\approx 0.02397\theta/D_\text{rod}$ is the maximum bending relaxation time. The black curve is for the rigid rod ($\theta=0$) for comparisons.}
\end{figure}

Similarly, the time correlation function $\langle \bm{P}(t)\cdot\bm{P}(0) \rangle$ can be calculate as follows:
\begin{align} \label{PtP0}
    \langle \bm{P}(t)\cdot\bm{P}(0) \rangle ={}& 
    4\langle \bm{N}(t)\cdot\bm{N}(0)\rangle +4\sum_{p=2,\text{even}}^\infty f_p(1) \big\langle \bm{N}(t)\cdot\bm{U}_p(0) \big\rangle \notag \\
    &+4\sum_{p=2,\text{even}}^\infty f_p(1)
    \big\langle \bm{N}(0)\cdot\bm{U}_p(t)\big\rangle
    + 4\sum_{p=2,\text{even}}^\infty f_p^2(1)\langle \bm{U}_p(t)\cdot\bm{U}_p(0)\rangle.
\end{align}
Using eq.~\eqref{eq:un} directly, we obtain
\begin{subequations} \label{89}
	\begin{align}
		\langle \bm{N}(t)\cdot\bm{N}(0)\rangle &= \left[\left(\frac{L}{2}\right)^2
                - \sum_{p=1}^\infty\sum_{q=1}^\infty \Gamma_{pq}\langle\bm{u}_p\cdot\bm{u}_q\rangle_\text{eq}\right]\langle\bm{n}(t)\cdot\bm{n}(0)\rangle+o(\theta), \\
		\langle \bm{N}(t)\cdot\bm{U}_p(0)\rangle &= \langle \bm{N}(0)\cdot\bm{U}_p(t)\rangle = \sum_{q=1}^\infty\sum_{r=1}^\infty \Gamma_{pqr}\langle\bm{u}_q\cdot\bm{u}_r\rangle_\text{eq}\langle \bm{n}(t)\cdot\bm{n}(0) \rangle+o(\theta), \label{NUparallel}
	\end{align}
\end{subequations}
where $\langle \bm{n}(t)\cdot\bm{u}_p(0)\rangle=\langle \bm{n}(0)\cdot\bm{u}_p(t)\rangle=0$ and eq.~\eqref{decoupled} has been used. As shown in eq.~\eqref{NUparallel}, the time-reversal symmetry is guaranteed for $p=1,2,3,\dots$.
Inserting eq.~\eqref{89} into eq.~\eqref{PtP0}, we obtain
\begin{align}
    \langle \bm{P}(t)\cdot\bm{P}(0) \rangle ={}& 4\left[\left(\frac{L}{2}\right)^2
                - \sum_{p=1}^\infty\sum_{q=1}^\infty \left(\frac{1}{2}\int_{-1}^1ds f_p'(s)f_q'(s)\right)\langle\bm{u}_p\cdot\bm{u}_q\rangle_\text{eq}\right]\langle\bm{n}(t)\cdot\bm{n}(0)\rangle \notag \\
    &+8\sum_{p=2,\text{even}}^\infty \langle \bm{u}_p(t)\cdot\bm{u}_p(0)\rangle  +o(\theta), \notag \\
    = {}& L^2\left[\left(1 - \alpha_R\theta\right)
    \exp \bigl(-2D_{\mathrm{r}}t \bigr) 
    + 2\theta\sum_{p=2,\text{even}}^\infty\frac{1}{\lambda_p^4} 
    \exp\biggl(-\frac{t}{\tau_p}\biggr)
    \right] +o(\theta). \label{ptp0res}
\end{align}
The second term is obtained by combining Eqs.~\eqref{equ:e42} for $\langle \bm{u}_p(t)\cdot\bm{u}_p(0)\rangle$ and \eqref{eq:solnt} for $\langle\bm{n}(t)\cdot\bm{n}(0)\rangle$, where $\alpha_R \approx 0.34286$ is given by eq.~\eqref{alphaR} in Appendix ``Eigen solutions.''
For the rigid-rod limit, $\left.\langle \bm{P}(t)\cdot\bm{P}(0) \rangle\right|_{B \to \infty} = L^2\exp(-2D_\text{rod}t)$ agrees with the rod theory \cite{1986_Doi}. 
The comparison between the semiflexible polymers and the rigid-rod limit is shown in \figrefp{fig:f2.0}{a}.
The difference for the long-time behavior is from the difference of the rotational diffusion coefficient which is affected by semiflexibility. 


The mean square displacement of $\bm{P}(t)$ is also obtained from eq.~\eqref{ptp0res},
\begin{align} \label{eqn:msdp}
 \big\langle [\bm{P}(t)-\bm{P}(0)]^2 \big\rangle &= 
 \big\langle \bm{P}^2(t) \big\rangle
+ \big\langle \bm{P}^2(0) \big\rangle
 -2\langle \bm{P}(t)\cdot\bm{P}(0) \rangle \notag \\
	&= 2L^2\left\{\left(1 - \alpha_R\theta\right)\left[1-
 \exp(-2D_{\mathrm{r}}t) \right]
 +2\theta\sum_{p=2,\text{even}}^\infty\frac{1}{\lambda_p^4}
 \left[1-\exp\biggl(-\frac{t}{\tau_p}\biggr)\right]\right\}+o(\theta).
\end{align}
For rigid rod, we can recover $\left. \big\langle [\bm{P}(t)-\bm{P}(0)]^2 \big\rangle\right|_{B \to \infty} = 2L^2\left[1-\exp(-2D_\text{rod}t) \right]$. The comparison is shown in \figref{fig:f2.0}(b). Besides the difference in the long-time behavior, the short-time behavior is also affected by the bending fluctuation.
For $t$ much smaller than the longest bending relaxation time
$\tau_1$, eq.~\eqref{eqn:msdp} is asymptotic to
\begin{equation} \label{eqn:msdp1}
\lim_{t \to 0}\langle [\bm{P}(t)-\bm{P}(0)]^2 \rangle \sim \int_0^\infty dp \frac{1}{p^4}\left[1-\exp\biggl(-\frac{p^4t}{\tau_1}\biggr)\right] \sim \left(\frac{t}{\tau_1}\right)^{3/4}
\end{equation}
where the relation $\tau_p \sim \lambda_p^{-4} \sim p^{-4}$ has been used. 
Therefore, the scaling exponent is changed from 1 to $3/4$.

\section{Viscoelasticity}
\subsection{Molecular expression for the stress tensor}

We now investigate the influence of flow field on the dynamics of polymers and calculate the viscoelastic function of dilute solution of semiflexible polymers.
The stress tensor of the system can be calculated by eq.~\eqref{eq:stress}. 
For the Rayleighian of eq.~\eqref{equ:e13A}, this gives the following stress tensor
\begin{align} \label{stress}
    \bm{\sigma} 
   &=\upsilon\int d\Omega\psi \left\{2\zeta_1D\frac{\partial \ln{\psi}}{\partial \Rc}\Rc + 2\zeta_1D\left(\bm{\delta}-\bm{n}\bm{n}\right)\cdot\frac{\partial \ln{\psi}}{\partial \bm{n}}\bm{n}+\frac{2}{3}\bm{N}^2\left(\zeta_1-\zeta_2\right)\bm{n}\cdot\bm{\kappa}\cdot\bm{n}\bm{n}\bm{n}\right.
   \notag \\
    &\quad \left.{}+\zeta_1\sum_{p=1}^\infty \left[2D\left(\bm{\delta}-\bm{n}\bm{n}\right)\cdot\frac{\partial\ln{\psi}}{\partial \bm{u}_p}\bm{u}_p+\frac{1}{\tau_p}\bm{u}_p\bm{u}_p +\frac{2}{3}\left(1-\frac{\zeta_2}{\zeta_1}\right)\bm{n}\cdot\bm{\kappa}\cdot\bm{n}\sum_{q=1}^\infty \Gamma_{pq}\bm{u}_p\bm{u}_q\right.\right. 
    \notag \\
    &\quad \left.\left.{}+\left(1-\frac{\zeta_2}{\zeta_1}\right) \bm{n}\cdot(\bm{\kappa}+\bm{\kappa}^T)\cdot\bm{u}_p\left(\bm{n}\bm{u}_p+\bm{u}_p\bm{n}\right)\right]-\frac{1}{3}\left(\zeta_1-\zeta_2\right)\frac{d\bm{N}^2}{dt}\bm{n}\bm{n} \right\}+o(\theta) 
    \notag \\
	&= 3\upsilon \kB T\langle\bm{n}\bm{n}\rangle +\frac{\upsilon\zeta^\perp L}{2}\sum_{p=1}^\infty\left(-2D\left(\bm{\delta}-\langle\bm{n}\bm{n}\rangle\right)+\frac{1}{\tau_p}\langle\bm{u}_p\bm{u}_p\rangle\right)+\frac{\upsilon\zeta^\parallel L}{6}\left(\bm{\kappa}+\bm{\kappa}^T\right):\bm{\mathcal{M}}+o(\theta),
\end{align}
where $\bm{\mathcal{M}}$ is a fourth moment defined by
\begin{align}
	\bm{\mathcal{M}} ={}& \left(\langle\bm{N}^2\rangle_\text{eq}-\frac{8D}{3}\frac{\zeta^\parallel}{\zeta^\perp}\sum_{p=1}^\infty\sum_{q=1}^\infty \Gamma_{pq}^2\tau_q\right)\langle\bm{n}\bm{n}\bm{n}\bm{n}\rangle \notag \\
	&+\sum_{p=1}^\infty\Big(\Gamma_{pp}\langle\bm{n}\bm{n}\bm{u}_p\bm{u}_p\rangle + \Gamma_{pp}\langle\bm{u}_p\bm{u}_p\bm{n}\bm{n}\rangle + 3\langle\bm{n}\bm{u}_p\bm{n}\bm{u}_p\rangle+3\langle\bm{u}_p\bm{n}\bm{u}_p\bm{n}\rangle\Big).
\end{align}
In eq.~\eqref{stress}, the first equality is obtained by using
eq.~\eqref{17.0}, and the second equality is obtained by integral by parts, where eq.~\eqref{11.b} and the same procedure as in eq.~\eqref{decoupled} are used. 
The isotropic term is ignored in eq.~\eqref{stress}.
The terms of summation over $p$ come from the bending of 
the polymer, and the other terms come from the rotational Brownian motion.

To complete the calculation, we need to determine $\langle\bm{n}\bm{n}\rangle$, $\langle\bm{u}_p\bm{u}_p\rangle$, and higher moment $\bm{\mathcal{M}}$ under the flow field.
We use eq.~\eqref{equ:e48.0} to obtain these averages.  
For the rotation, we have
\begin{align}\label{eq:e42}
    \frac{\partial}{\partial t}\langle \bm{n}\bm{n} \rangle ={}& \langle\mathcal{L}^\dag \left(\bm{n}\bm{n}\right)\rangle  \notag \\
    ={}& -18D\bigg\langle \frac{1}{\bm{N}^2}\left(\bm{n}\bm{n} -\frac{1}{3}\bm{\delta}\right)\bigg\rangle+\bm{\kappa}\cdot\langle \bm{n}\bm{n} \rangle+\langle \bm{n}\bm{n} \rangle\cdot\bm{\kappa}^T-2\bm{\kappa}:\langle \bm{n}\bm{n}\bm{n}\bm{n} \rangle \notag \\
    ={}& -\frac{18D}{\langle\bm{N}^2\rangle_\text{eq}}\left(\langle \bm{n}\bm{n} \rangle-\frac{1}{3}\bm{\delta}\right)+\bm{\kappa}\cdot\langle \bm{n}\bm{n} \rangle+\langle \bm{n}\bm{n} \rangle\cdot\bm{\kappa}^T-2\bm{\kappa}:\langle \bm{n}\bm{n}\bm{n}\bm{n} \rangle+o(\theta),
\end{align}
where the same procedure as in eq.~\eqref{decoupled} has been used in the second equality up to the order of $\theta$.

For the bending deflections, we have
\begin{align} \label{eq:e43}
    \frac{\partial}{\partial t}\langle \bm{u}_p\bm{u}_p \rangle ={}& \langle\mathcal{L}^\dag \left(\bm{u}_p\bm{u}_p\right)\rangle  \notag \\
    ={}& 4D\left(\bm{\delta}-\langle\bm{n}\bm{n}\rangle\right)-\frac{2}{\tau_p}\langle \bm{u}_p\bm{u}_p \rangle+\bm{\kappa}\cdot\langle \bm{u}_p\bm{u}_p \rangle+\langle \bm{u}_p\bm{u}_p \rangle\cdot\bm{\kappa}^T \notag \\
    &-\bm{\kappa}:\langle\bm{n}\bm{u}_p\bm{n}\bm{u}_p\rangle-\langle\bm{u}_p\bm{n}\bm{u}_p\bm{n}\rangle:\bm{\kappa}^T 
    -\frac{4}{3}\frac{\zeta^\parallel}{\zeta^\perp}\Gamma_{pp}\bm{\kappa}:\langle\bm{n}\bm{n}\bm{u}_p\bm{u}_p\rangle.
\end{align}

Eqs.~\eqref{eq:e42} and \eqref{eq:e43} cannot be solved since
there are unknown fourth moments $\langle\bm{n}\bm{n}\bm{n}\bm{n}\rangle$ and $\langle\bm{n}\bm{n}\bm{u}_p\bm{u}_p\rangle$  in the equation. 
However, in the linear viscoelasticity where only the
first order perturbation by $\bm{\kappa}$ is accounted for, 
the fourth moments in \eqref{stress}--\eqref{eq:e43} can be replaced by its equilibrium value.  We will focus on this linear viscoelasticity in the following section, where the system undergoes a small velocity gradient $\bm{\kappa}$ with incompressible condition $\text{tr}(\bm{\kappa})=0$. 
In this case, eqs.~\eqref{eq:e42} and \eqref{eq:e43} can be solved analytically.

\subsection{The linear viscoelasticity}


The equilibrium value of the fourth moment is calculated as
\begin{equation} \label{eq:nnnn}
    \langle n_\alpha n_\beta n_\mu n_\nu \rangle_\text{eq} = \frac{1}{15}\left(\delta_{\alpha\beta}\delta_{\mu\nu}+\delta_{\alpha\mu}\delta_{\beta\nu}+\delta_{\alpha\nu}\delta_{\beta\mu}\right).
\end{equation}
The first-order solution of eq.~\eqref{eq:e42} for the rotation is then given by
\begin{equation} \label{equ:e48}
	\langle \bm{n}\bm{n} \rangle = \frac{1}{3}\bm{\delta}+\frac{1}{5}\int_{-\infty}^tdt'
 \exp\biggl(-6D_{\mathrm{r}}(t-t')\biggr) 
 \left[\bm{\kappa}(t')+\bm{\kappa}^T(t')\right].
\end{equation}
The solution for $\langle \bm{n}\bm{n} \rangle$ is consistent with the rigid rod but with the modified rotational diffusion coefficient.

Also, we have the equilibrium quantity
\begin{equation} \label{eq44.0}
	\langle n_\alpha n_\beta u_{p,\mu} u_{q,\nu} \rangle_\text{eq} = \frac{2}{15}D\tau_p\delta_{pq}\left(4\delta_{\alpha\beta}\delta_{\mu\nu}-\delta_{\alpha\mu}\delta_{\beta\nu}-\delta_{\alpha\nu}\delta_{\beta\mu}\right)+o(\theta),
\end{equation}
which can be obtained by solving equation $\big\langle\mathcal{L}^\dag_0 \left(\bm{n}\bm{n}\bm{u}_p\bm{u}_q\right)\big\rangle=0$. The first-order approximate solution of eq.~\eqref{eq:e43} for bending is
\begin{align}
    \langle \bm{u}_p\bm{u}_p \rangle ={}& 2D\tau_p\big(\bm{\delta}-\langle\bm{n}\bm{n}\rangle\big) + \left(\frac{14}{15}+\frac{8}{45}\frac{\zeta^\parallel}{\zeta^\perp}\Gamma_{pp}\right)D\tau_p\int_{-\infty}^tdt'
    \exp\biggl(-\frac{2}{\tau_p}(t-t')
    \biggr)
    \left[\bm{\kappa}(t')+\bm{\kappa}^T(t')\right], \label{eq:68.0}
\end{align}
where the similar Laplace transform procedure as in eq.~\eqref{decoupled} has been used. 
Note that the rotation term will cancel with that in stress expression \eqref{stress}. 
The result given by the trace of eq.~\eqref{eq:68.0} is also consistent with the equilibrium values in eq.~\eqref{eqvalue}.

Using these result,  we have
\begin{equation}
    \bm{\sigma} = 
    \int_{-\infty}^tdt' G(t-t')\left[\bm{\kappa}(t')
    +\bm{\kappa}^T(t')\right],
\end{equation}
where the relaxation modulus is given by
\begin{equation}
  G(t) = \frac{3}{5}\upsilon \kB T
 \exp(-6D_{\mathrm{r}}t)
 +\upsilon \kB T\sum_{p=1}^\infty G_p
 \exp\biggl(-\frac{2}{\tau_p}t \biggr) 
 + \upsilon \kB T \frac{G_0}{6D_{\mathrm{r}}}\delta(t) + o(\theta),
\end{equation}
with $D_{\mathrm{r}}$ given by eq.~\eqref{DiffsionC}, $\tau_p=(L/2)^3\zeta^\perp/B\lambda_p^4$ ($p=1,2,3,\dots$), and
\begin{subequations}
	\begin{align}
		G_p &= \frac{7}{15}+\frac{4}{45}\frac{\zeta^\parallel}{\zeta^\perp}\Gamma_{pp} \sim p^2, \label{Gp.eq}\\
		G_0 &= \frac{2}{5}\frac{\zeta^\parallel}{\zeta^\perp}\left[1-\left(\alpha_G+\frac{\zeta^\parallel}{\zeta^\perp}\alpha_{[a]}\right)\theta\right]. \label{G0.eq}
	\end{align}
\end{subequations}

\begin{figure}[htb]
\centering
\includegraphics[width=0.9\textwidth]{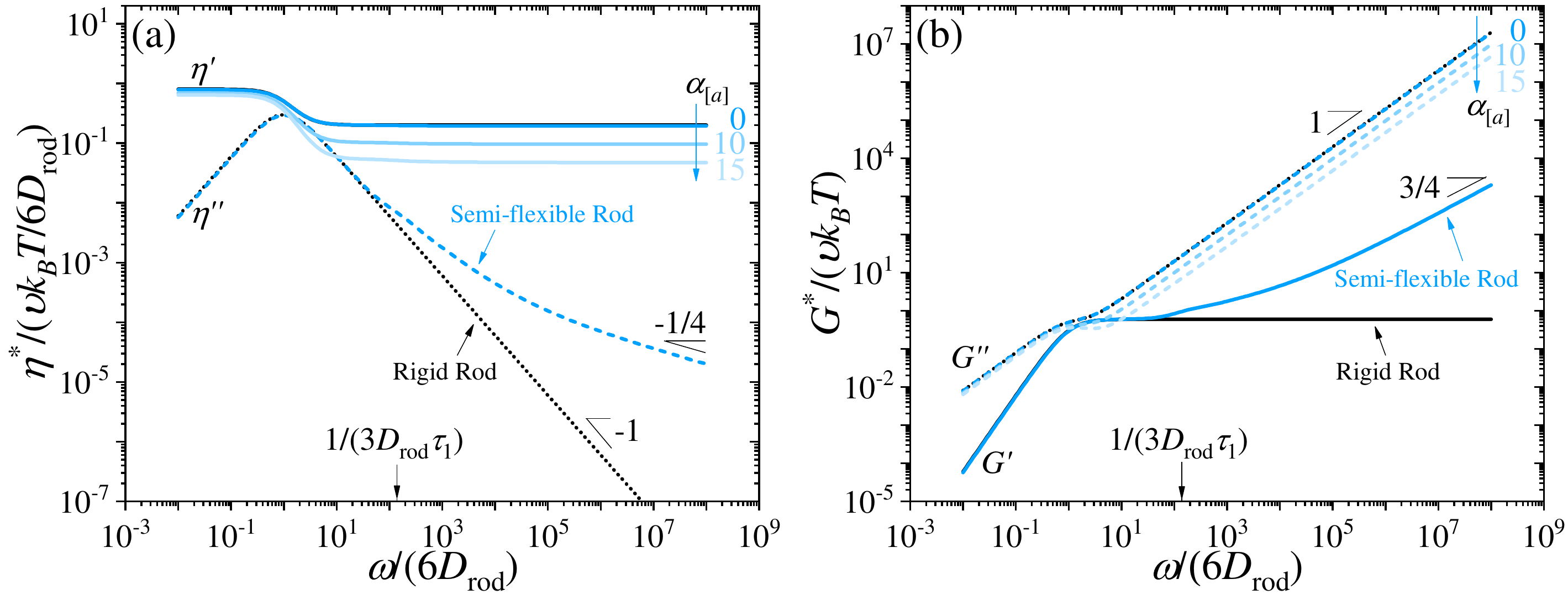}
\caption{\label{fig:f2x}
Typical behaviors of (a) the dynamic viscosities $\eta'(\omega)$ and $\eta''(\omega)$ and (b) the dynamic moduli $G'(\omega)$ and $G''(\omega)$ for $\theta=1/10$ (blue curve). 
The black curve is for the rigid rod ($\theta=0$) for comparisons. 
The material-dependent constant $\alpha_{[a]}=0$, $10$, or $15$ 
has been used for example, which only shifts the high-frequency curve of $\eta'$ and $G''$.}
\end{figure}

In eq.~\eqref{Gp.eq} for $p \gg 1$, there is an asymptotic relation from eq.~\eqref{asymptotic}. 
In eq.~\eqref{G0.eq}, $\alpha_G $ is a constant defined by eq.~\eqref{alphaG} in Appendix ``Eigen solutions'' ($\alpha_G\approx 0.08571$), and $\alpha_{[a]}$ is defined by
\begin{equation}
    \alpha_{[a]} = \frac{2}{3}\sum_{p=1}^{\infty}\sum_{q=1}^{\infty} \frac{\Gamma_{pq}^2}{\lambda_q^4}.
\end{equation}
Notice that the summation on the right-hand side diverges since for large $p$, both $\Gamma_{pp}^2$ and $\lambda_p^4$ are asymptotic to $p^4$. 
The divergence comes from $d\bm{N}^2/dt$ in stress \eqref{stress}, which goes to infinity as $\omega$ goes to infinity.  However, the inextensible condition we are using will not be realistic for infinitely large $\omega$ (since the sound speed is finite). To avoid the divergence, we need to consider the characteristic length below which the continuum description (or the inextensible condition) cannot be used.
For such a semiflexible polymer, we suppose that the wavelength of the bending is limited to the finite thickness $a$, and $p$ and $q$ must be less than $L/a$.
The finite thickness of the polymer is also needed to have a finite value of $\zeta^\perp$ (or $\zeta^\parallel$)\cite{1986_Doi}.
Therefore, we need to calculate $\alpha_{[a]}$ by the following equation:
\begin{equation}
    \alpha_{[a]} 
    = \frac{2}{3}\sum_{p=1}^{L/a}\sum_{q=1}^{L/a} \frac{\Gamma_{pq}^2}{\lambda_q^4} \sim  \sum_{p=1}^{L/a}\sum_{q=1}^{L/a} \frac{p^2}{q^2} 
    \sim \left(\frac{L}{a}\right)^3.
\end{equation}
This equation indicates that $\alpha_{[a]}$ is large.
On the other hand, $\alpha_{[a]}$ cannot be very large since $G_0$ in eq.~\eqref{G0.eq} must be positive. 
Therefore, $G_0$ in eq.~\eqref{G0.eq} cannot be determined uniquely by the present theory. 
This is not a failure of the present theory; it is a consequence that $G_0$ represents the response at infinitely short time and cannot be described by the present continuum model for the chain. 
%
%
Therefore, $\alpha_{[a]}$ is actually a number determined by the material properties of the semiflexible rod itself. Subsequently, we will take $\alpha_{[a]}=0,10,15$ for example (see \figref{fig:f2x}).

From the relaxation modulus $G(t)$, we can calculate the
the complex viscosity by $\eta^*(\omega)=\int_0^\infty dt\exp(- i \omega t)G(t)$ as
\begin{align}
    \eta^*(\omega)&=\frac{3}{5}\upsilon \kB T\frac{1/6D_{\mathrm{r}}}{1+i\omega/6D_{\mathrm{r}}}
    +\upsilon \kB T\sum_{p=1}^\infty G_p\frac{\tau_p/2}{1+i \omega\tau_p/2}+\upsilon \kB T \frac{G_0}{6D_{\mathrm{r}}}.     \label{eqn:73}
\end{align}

At high and low frequencies, it approaches to:
\begin{subequations}
    \begin{align}
        \lim_{\omega \to \infty}\eta^*(\omega) &= \upsilon \kB T \frac{G_0}{6D_{\mathrm{r}}},   \label{eqn:74a}\\
        \lim_{\omega \to 0}\eta^*(\omega) &= \upsilon \kB T \frac{G_0}{6D_{\mathrm{r}}} + \upsilon \kB T\frac{1}{10D_{\mathrm{r}}} +\frac{1}{2}\upsilon \kB T\sum_{p=1}^\infty G_p\tau_p.   \label{eqn:74b}
    \end{align}
\end{subequations}
In the rigid-rod limit of $B \to \infty$ ($\theta\to 0$ and $\tau_p \to 0$),  one recovers the result of rigid rod \cite{1986_Doi} $\eta^*(0)/\eta^*(\infty) = 4$ using the relation $\zeta^\perp=2\zeta^\parallel$.

The dynamic moduli $G^*(\omega) = G'(\omega)+i G''(\omega) = i \omega\eta^*(\omega)$ is given by
\begin{subequations}
	\begin{align}
		G'(\omega) &= \frac{3}{5}\upsilon \kB T\frac{\left(\omega/6D_{\mathrm{r}}\right)^2}{1+\left(\omega/6D_{\mathrm{r}}\right)^2}+\upsilon \kB T\sum_{p=1}^\infty G_p\frac{\left(\omega\tau_p/2\right)^2}{1+\left(\omega\tau_p/2\right)^2}, \\
		G''(\omega) &= \frac{3}{5}\upsilon \kB T\frac{\omega/6D_{\mathrm{r}}}{1+\left(\omega/6D_{\mathrm{r}}\right)^2}+\upsilon \kB T\sum_{p=1}^\infty G_p\frac{\omega\tau_p/2}{1+\left(\omega\tau_p/2\right)^2}+G_0\upsilon \kB T\omega/6D_{\mathrm{r}}.
	\end{align}
\end{subequations}

The typical behaviors of dynamic viscosity and dynamic moduli are shown in \figref{fig:f2x}. 
It is seen that the difference between the semiflexible polymer and the rigid rod is very small at low frequencies, but becomes conspicuous at high frequency: the storage modulus $G'(\omega)$ of rigid rod approaches to a constant value while $G'(\omega)$ of semiflexible polymer keeps increasing with $\omega$ following the scaling relation 
$G'(\omega) \sim \omega^{3/4}$. 
We can show this asymptotic behavior of semiflexible polymer using the relation $G_p \sim p^2$ and $\tau_p \sim p^{-4}$, as
\begin{equation}
    \lim_{\omega \to \infty}G'(\omega) \sim \sum_{p=1}^\infty G_p\frac{\left(\omega\tau_p\right)^2}{1+\left(\omega\tau_p\right)^2} \sim \int_0^\infty dp \frac{p^2\left(\omega\tau_1\right)^2}{p^8+\left(\omega\tau_1\right)^2} \sim \left(\omega\tau_1\right)^{3/4}.
\end{equation}
This scaling relation has been shown by previous works \cite{1996_Amblard,2002_Shankar}.

On the other hand, the loss modulus $G''(\omega)$ at high frequency follows the scaling relation $G''(\omega) \sim \omega$, but its magnitude depends on the value of $\alpha_{[a]}$. 
Similarly, the real part of the complex viscosity $\eta'(\omega)$ approaches to a constant value which depends on the parameter $\alpha_{[a]}$ as well.

\section{Conclusions}
In this paper, we have developed a systematic perturbation 
calculation from the rod theory. 
Using the orthonormal eigenfunctions, 
we defined the normal coordinates for the
translational mode, the rotational mode, and the bending modes.
The complexity arises from the inextensibility constraint, which causes the coupling between the rotational mode and the bending modes. 
By regarding the bending as a small quantity of
the order of $\theta \ll 1$, we constructed the Rayleighian of the system and derived the Smoluchowski equation for the distribution function of polymer conformation. 
We then calculated the time correlation functions, which characterize the Brownian motion at equilibrium, and the response functions for the linear viscoelasticity.
The present work predicts the same equilibrium value of mean square end-to-end distance as the classical WLC model. 
The equilibrium time correlation function of the end-to-end vector follows the typical scaling law $t^{3/4}$ in the early stage, which corresponds to the bending fluctuations. 
The scaling law $\omega^{3/4}$ of dynamic moduli in the high-frequency limit of semiflexible linear polymers agrees with the typical behavior of semiflexible linear polymers as well.
The rotational diffusion coefficient increases slightly by semiflexibility because the equilibrium end-to-end length of the semiflexible polymer is slightly smaller than that of the rigid rod with the same contour length.
The high-frequency viscosity is shown to be a quantity which
depends on the discrete molecular structure of the semiflexible polymers, where polymers cannot be regarded as a curve of zero thickness.

\begin{acknowledgement}

The authors thank the support of the National Natural Science Foundation of China (NSFC) (Nos.\,12247174, 12174390, and 12150610463) and Wenzhou Institute, University of Chinese Academy of Sciences (WIUCASQD2022004 and WIUCASQD2020002), and Oujiang Laboratory (OJQDSP2022018).

\end{acknowledgement}

\section{Appendix} 
\subsection{Eigen solutions}
\label{Appendix:EigenSolutions}

\begin{figure}[htb]
\centering
\includegraphics[width=0.6\textwidth]{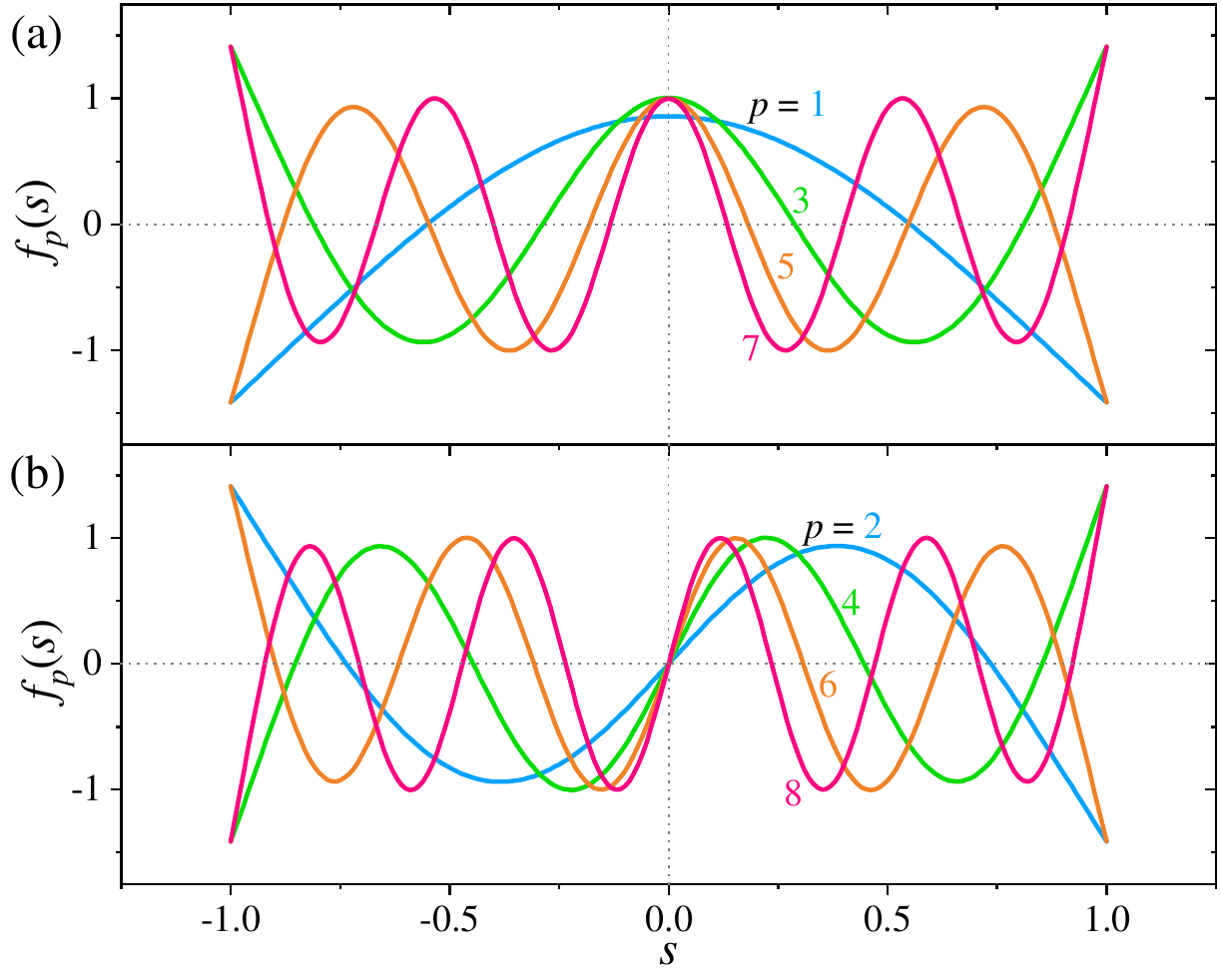}
\caption{\label{fig:f1}
The first eight bending modes: 
(a) first four even eigen modes and (b) first four odd eigen modes, respectively.}
\end{figure}

The eigenvalues and the eigenfunctions defined by the eigenvalue problem\cite{1985_Aragon,1998_Wiggins,2002_Shankar,2007_Young,2012_Kantsler}
eqs.~\eqref{equ:e1} and \eqref{equ:e2} in $s\in[-1,1]$ are: 

\begin{subequations} 
\label{Eigen solutions}
\noindent
$\bullet$ the even parts
\begin{equation}
    \begin{cases}
        \tan{(\lambda_p)}+\tanh{(\lambda_p)}=0, \\[2ex]
        f_p(s)=\frac{1}{\sqrt{\cosh{^2(\lambda_p)}+\cos{^2(\lambda_p)}}}\left[\cos{(\lambda_p)}\cosh{(\lambda_p s)}+\cosh{(\lambda_p)}\cos{(\lambda_p s)}\right], 
    \end{cases}
\end{equation}
where the asymptotic solutions of the eigenvalues are $\lambda_p\approx (2p+1)\pi/4$ with $p=1,3,5,\dots$.

\noindent
$\bullet$ the odd parts
\begin{equation}
    \begin{cases}
        \tan{(\lambda_p)}-\tanh{(\lambda_p)}=0,  \\[2ex]
        f_p(s)=\frac{1}{\sqrt{\sinh{^2(\lambda_p)}-\sin{^2(\lambda_p)}}}\left[\sin{(\lambda_p)}\sinh{(\lambda_ps)}+\sinh{(\lambda_p)}\sin{(\lambda_p s)}\right], 
    \end{cases}
\end{equation}
where the asymptotic solutions of the eigenvalues are $\lambda_p\approx (2p+1)\pi/4$ with $p=2,4,6,\dots$. 


The first eight bending modes are shown in \figref{fig:f1}.
\end{subequations}

Three properties for the eigenfunctions are
\begin{subequations}\label{eq.f3}
\begin{gather} 
   \int_{-1}^1d sf'''_p(s)=f''_p(s)\Big|_{-1}^1 = 0, 
   \\\int_{-1}^1dsf'''_p(s)f'_q(s)=f'''_p(s)f_q(s)\Big|_{-1}^1-\int_{-1}^1dsf''''_p(s)f_q(s)=-\lambda_p^4\delta_{pq}, 
   \\\int_{-1}^1dsf''_p(s)f''_q(s)=f''_p(s)f'_q(s)\Big|_{-1}^1-\int_{-1}^1dsf'''_p(s)f'_q(s)=\lambda_p^4\delta_{pq}, \label{f2f2}
\end{gather}
\end{subequations}
where the boundary conditions \eqref{equ:e2} and the orthogonality \eqref{equ:e1A} of the eigenfunctions are used.

For $p,q \gg 1$, one has the asymptotic relations for the integration with any continue function $g(s)$:
\begin{subequations}
    \begin{gather}
        \Gamma_{pp} = \int_{-1}^1ds g(s)f_p^{\prime 2}(s) = g(\xi_1) \int_{-1}^1ds f_p^{\prime 2}(s) \sim \lambda_p^2 \sim p^2, \label{asymptotic} \\
        \Gamma_{pq}^2=\left(\int_{-1}^1ds g(s)f_p'(s)f_q'(s)\right)^2 \leq g^2(\xi_2)\int_{-1}^1ds f_p^{\prime 2}(s)\int_{-1}^1ds f_q^{\prime 2}(s) \sim p^2q^2, \label{Cpq2}
\end{gather}
\end{subequations}
where existing at least one point $\exists \xi_1,\xi_2 \in [-1,1]$ satisfies the second equality in eq.~\eqref{asymptotic} by the mean value theorem for definite integrals since $f_p^{\prime 2}(s)$ is a non-negative continue function.
The integral is estimated based on the explicit expression for $f_p(s)$ in eq.~\eqref{Eigen solutions}, which is found to be asymptotic to $\lambda_p^2$. 
Additionally, Cauchy-Buniakowsky-Schwarz inequality is used to obtain eq.~\eqref{Cpq2}.

One can also prove that $f_p^2(1)=2$ for $p=1,2,3,\dots$ by using $\tan^2{(\lambda_p)}=\tanh^2{(\lambda_p)}$, where
\begin{equation}
	f_p(1) =
	\begin{cases}
        (-1)^{(p+1)/2}\sqrt{2}, & p=1,3,5,\dots,  \\[2ex]
        (-1)^{p/2}\sqrt{2}, & p=2,4,6,\dots,
    \end{cases}
\end{equation}
and calculate some numerical coefficients
\begin{subequations}
	\begin{align}
		\alpha_R &= \frac{1}{2}\sum_{p=1}^\infty \frac{1}{\lambda_p^4}\int_{-1}^1ds f_p'^2(s) \approx 0.34286, \label{alphaR} \\
	\alpha_P &= \alpha_R-2\sum_{r=2,\text{even}}^\infty \frac{1}{\lambda_r^4} \approx 0.33333, \label{alphaP} \\
	\alpha_D &= \sum_{p=1}^\infty\frac{\Gamma_{pp}}{\lambda_p^4}= \alpha_R+\sqrt{2}\sum_{p=1}^\infty\sum_{r=2,\text{even}}^\infty \frac{(-1)^{r/2}}{\lambda_r^4\lambda_p^4}\int_{-1}^1dsf'''_r(s)f_p^{\prime 2}(s) \approx 0.25714, \label{alphaD} \\
	\alpha_G &= \alpha_D-\frac{9}{2}\sum_{p=1}^\infty\frac{1}{\lambda_p^4} \approx 0.08571. \label{alphaG}
	\end{align}
\end{subequations}

\subsection{Inextensibility}
Inserting eq.~\eqref{equ:e4a} into eq.~\eqref{eqn.13}, one has
\begin{equation}
    \bm{N}^2+2\sum_{p=1}^\infty \bm{N}\cdot\bm{U}_p f_p'(s)+\sum_{p=1}^\infty\sum_{q=1}^\infty\bm{U}_p\cdot\bm{U}_qf_p'(s)f_q'(s) = \left(\frac{L}{2}\right)^2. \label{e.130}
\end{equation}

Integrating both sides of eq.~\eqref{e.130} by $\int_{-1}^1dsf'''_r(s)$ and using eq.~\eqref{eq.f3}, we get
\begin{equation}
	\bm{N}\cdot\bm{U}_r = \sum_{p=1}^\infty\sum_{q=1}^\infty \Gamma_{rpq}\bm{U}_p\cdot\bm{U}_q, \label{eq:15a.0}
\end{equation}
with
\begin{equation}
    \Gamma_{rpq}=\frac{1}{2\lambda_r^4}\int_{-1}^1dsf'''_r(s)f_p'(s)f_q'(s),
    \label{eq:15a.0A}
\end{equation}
where $r,p,q=1,2,3,\dots$. 
Similarly, integrating both sides of eq.~\eqref{e.130} by $\int_{-1}^1ds$ and using eq.~\eqref{eq:15a.0}, we get
\begin{equation}
    \bm{N}^2 = \left(\frac{L}{2}\right)^2 -  \sum_{p=1}^\infty\sum_{q=1}^\infty \Gamma_{pq}\bm{U}_p\cdot\bm{U}_q, \label{eq:16.0}
\end{equation}
with
\begin{equation} \label{Cpq}
    \Gamma_{pq}=2\sqrt{2}\sum_{r=2,\text{even}}^\infty (-1)^{r/2}\Gamma_{rpq}+\frac{1}{2}\int_{-1}^1ds f_p'(s)f_q'(s),
\end{equation}
where $p,q=1,2,3,\dots$.

\subsection{Constraint conditions}
Using the chain rule of derivatives for a transformation $\bm{y}=\bm{M}^T\cdot\bm{x}$, one has\cite{1987_Bird}
\begin{equation}
    \frac{\partial }{\partial\bm{x}}=\bm{M}\cdot\frac{\partial }{\partial\bm{y}}.
\end{equation}

Specifying that $\bm{M}=\bm{\delta}-\bm{n}\bm{n}$ and $\bm{x}=\bm{u}_p$, one has $\bm{y}=\bm{u}_p$ since there is the constraint eq.~\eqref{eq:15.0b}. Therefore,
\begin{equation} \label{eq:65pd}
    \bm{n}\cdot\frac{\partial \psi}{\partial\bm{u}_p} = \bm{n}\cdot\left(\bm{\delta}-\bm{n}\bm{n}\right)\cdot\frac{\partial \psi}{\partial\bm{u}_p} = 0,
\end{equation}
which means that $\bm{u}_p$ are essentially two dimensional vectors perpendicular to $\bm{n}$ and have properties
\begin{equation}
    \frac{\partial }{\partial\bm{u}_p}\bm{u}_p = \bm{\delta}-\bm{n}\bm{n}, \qquad
    \frac{\partial }{\partial\bm{u}_p}\cdot\bm{u}_p = 2.
\end{equation}

Similarly, specifying that $\bm{M}=\bm{\delta}-\bm{u}_p\bm{u}_p/\bm{u}_p^2$ and $\bm{x}=\bm{n}$, we have
\begin{equation}\label{eq:66pd}
    \bm{u}_p\cdot\frac{\partial \psi}{\partial\bm{n}} = \bm{u}_p\cdot\left(\bm{\delta}-\bm{u}_p\bm{u}_p/\bm{u}_p^2\right)\cdot\frac{\partial \psi}{\partial\bm{n}} = 0.
\end{equation}

%
%

\bibliography{ref}

\end{document}